\documentclass[10pt,journal]{IEEEtran}
\IEEEoverridecommandlockouts

\usepackage{cite}
\usepackage{amsmath,amssymb,amsfonts}
\usepackage{algorithmic}
\usepackage{graphicx}
\usepackage{textcomp}
\usepackage{xcolor}
\usepackage{makecell} 
\usepackage{multirow}
\usepackage{calrsfs} 
\usepackage{url} 
\usepackage{pifont}
\newcommand{\xmark}{\ding{55}}%

\def\BibTeX{{\rm B\kern-.05em{\sc i\kern-.025em b}\kern-.08em
    T\kern-.1667em\lower.7ex\hbox{E}\kern-.125emX}}
    
\begin{document}
\title{\huge Surviving the Storm: The Impacts of Open RAN Disaggregation on Latency and Resilience}

\author{
\IEEEauthorblockN{
    Sotiris Chatzimiltis\IEEEauthorrefmark{1}, \textit{Graduate Student Member, IEEE}, 
    Mohammad Shojafar\IEEEauthorrefmark{1}, \textit{Senior Member, IEEE}, \\
    Mahdi Boloursaz Mashhadi\IEEEauthorrefmark{1}, \textit{Senior Member, IEEE}, 
    and Rahim Tafazolli\IEEEauthorrefmark{1}, \textit{Fellow, IEEE}
}\\
\IEEEauthorblockA{
    \IEEEauthorrefmark{1}5G/6GIC, Institute for Communication Systems (ICS), University of Surrey, Guildford, UK \\
    \{sc02449, m.shojafar, m.boloursazmashhadi, r.tafazolli\}@surrey.ac.uk
}
}
\maketitle

\begin{abstract}
The development of Open Radio Access Networks (Open RAN), with their disaggregated architectures and virtualization of network functions, has brought considerable flexibility and cost savings to mobile networks. However, these architectural advancements introduce additional latency during the initial attachment procedure of User Equipment (UE), increasing the risk of signaling storms. This paper investigates the latency impact due to disaggregation of the Base-band Unit (BBU) into the Central Unit (CU) and Distributed Unit (DU). Specifically, we model the delays induced due to disaggregation on UE attachment, analyzing the performance under varying load conditions, and sensitivity to processing times. We demonstrate that while both monolithic and Open RAN architectures experience performance degradation under high-load conditions, Open RAN's added overheads can increase its susceptibility to congestion and signaling storms.
However, Open RAN's inherent flexibility, enabled by disaggregation and virtualization, allows efficient deployment of resources, faster service deployment, and adaptive congestion control mechanisms to mitigate these risks and enhance overall system resilience. Thereby, we quantify resilience by introducing a new utility function and propose a novel adaptation mechanism to reinforce Open RAN's robustness against signaling storms. Our results show that the proposed adaptive mechanism significantly enhances resilience, achieving improvements of up to 286\% over fixed configurations, with resilience scores approaching 0.96 under optimal conditions. While simulation results show that Open RAN disaggregation increases attachment latency and susceptibility to signaling congestion, they also highlight that its architectural flexibility can mitigate these effects, improving resilience under high-load conditions.

\end{abstract}


\section{Introduction}
The rapid evolution of mobile communication networks, driven by the demand for higher performance, flexibility, and cost efficiency, has led to the emergence of Open Radio Access Networks (Open RAN). Unlike traditional monolithic radio access networks, which rely on proprietary, tightly integrated hardware and software, Open RAN promotes a disaggregated architecture where the Centralized Unit (CU), Distributed Unit (DU), and Radio Unit (RU) are decoupled and can be provided by different vendors. This disaggregation, combined with the virtualization of the units as network functions, is designed to foster interoperability, scalability, and dynamic resource allocation. However, these architectural innovations also introduce new challenges, such as delays and overheads.

One of the critical procedures in any mobile network is the User Equipment (UE) initial attachment, where signaling plays a fundamental role in establishing communication between the UE and the network. In traditional RAN architectures, this process is relatively simplified, as all components responsible for managing signaling are integrated within a single base station. In contrast, Open RAN introduces additional delays as signaling traffic must traverse multiple disaggregated components and interfaces, such as the F1 interface between the DU and CU and the open fronthaul (O-FH) interface between the DU and the RU. 

The flexibility provided by disaggregation and virtualization can be advantageous for network scalability and vendor diversity, enabling operators to adopt a multi-vendor approach and tailor network deployments to specific needs. This adaptability not only allows innovation but also cost optimization and the efficient use of resources. However, it raises significant concerns about handling large volumes of signaling traffic, particularly during high-demand scenarios like network recovery or high-density events, due to the additional overheads introduced. A key concern is the potential for signaling storms, a phenomenon where excessive signaling messages overwhelm the network's control plane, leading to service degradation or complete outages. In Open RAN, the increased complexity and delay in handling signaling messages may worsen this issue, making the system more vulnerable to signaling storms compared to monolithic RAN architectures. Ensuring resilience in Open RAN during high signaling loads is essential to safeguard both network availability (the ability of the network to remain accessible and operational) and reliability (consistent performance under adverse conditions). Resilience in Open RAN goes beyond traditional reliability and robustness by integrating adaptive mechanisms for real-time detection, response, and recovery, ensuring the network can reorganize and maintain functionality during unforeseen challenges. Embedding these principles is critical to mitigate the impact of disruptions such as signaling storms, which can arise from malicious attacks or high-density user scenarios, and to enable swift recovery with minimal impact on service quality.

This paper aims to mathematically model the impact of disaggregation on the UE attachment procedure in Open RAN and compare it with traditional RAN. By identifying the additional delays introduced by Open RAN's architecture, we explore how these delays contribute to signaling congestion and the conditions under which signaling storms may arise. The results from this analysis will provide insights into the trade-offs between the flexibility of Open RAN and its potential vulnerability to signaling storms.

\subsection{Contributions}
The contributions of this paper are as follows:
\begin{itemize}
    \item \textbf{Comprehensive Analysis of the F1 Interface and RRC Protocols:}  
    We provide a detailed breakdown of the F1 Application Protocol (F1AP) and the Radio Resource Control (RRC) protocol, focusing on their information elements (IEs) and their impact on Open RAN signaling. This analysis serves as a foundation for understanding how these protocols influence system performance and delay accumulation in disaggregated architectures.

    \item \textbf{Characterization of Delay Components in the F1 Interface:}  
    We analyze the delays introduced by the F1 interface in Open RAN, considering its role in control-plane signaling between the CU and DU. Our study quantifies key delay components, including transmission, propagation, queuing, and processing delays, highlighting their impact on the UE initial attachment procedure. 
    
    \item \textbf{Performance Evaluation under High Load and Signaling Storm Conditions:}  
    We conduct a comparative analysis of monolithic and Open RAN architectures, examining variations in service rates, system utilization, and latency across different scenarios. Additionally, we simulate a mass UE attachment event where the arrival rate exceeds the service rate, demonstrating the impact on system congestion and delay accumulation.

    \item \textbf{Development of a Utility Function and Adaptive Resilience Mechanism:}  
    To assess system resilience under dynamic traffic conditions, we introduce a novel utility function that quantifies the network's ability to absorb, adapt, and recover from signaling storms. Furthermore, we propose an adaptive service mechanism that dynamically adjusts resource allocation to enhance Open RAN resilience, mitigating the effects of congestion and improving recovery times.
\end{itemize}

\subsection{Organization}
The remainder of this paper is structured as follows. Section~\ref{sec:rel_work} reviews related work on Open RAN disaggregation, signaling storm attacks, and network resilience. Section~\ref{sec:openran} provides an overview of the Open RAN architecture, detailing its components, protocol stacks, and key interfaces. Section~\ref{sec:ueia} presents an in-depth analysis of the UE initial attachment procedure, beginning with delay modeling (\ref{ssec:md}), followed by delay accumulation and performance evaluation under varying load conditions (\ref{ssec:da_pe}). Additionally, we analyze the impact of a mass UE attachment scenario (\ref{ssec:massue}) on network congestion. The section concludes with the introduction of a quantitative resilience metric and a utility function to assess system performance under different conditions (\ref{ssec:res}). Section~\ref{sec:qaad} evaluates system behavior, analyzing delays, performance metrics, utility function values under both normal operation and signaling storm conditions, and provides resilience scores for several scenarios. Finally, Section~\ref{sec:concl} summarizes the findings and outlines potential future research directions.

\begin{table}[!t]
\centering
\caption{Description of notation used}
\label{tab:notation}
\resizebox{\columnwidth}{!}{%
\begin{tabular}{ll}
\hline
\textbf{Symbol} & \textbf{Description} \\\hline
$c$ & Number of servers \\
$c_{\max}$ & Maximum number of servers available \\
$D_p$ & Propagation delay \\
$D_q$ & Queuing delay \\
$D_r$ & Processing delay \\
$D_t$ & Transmission delay \\
$D_{\text{RU-BBU}}$ & One-way delay between RU and BBU in monolithic RAN \\
$D_{\text{RU-CU}}$ & One-way delay from RU to CU (via DU) in Open RAN \\
$D_{\text{total}}$ & Total delay \\
$k_A, k_B$ & Steepness factors in the utility function \\
$\lambda$ & Arrival rate (UEs/sec) \\
$L(t)$ & Lyapunov function (measuring system stability) \\
$L_q$ & Queue length \\
$L_{\text{qmax}}$ & Maximum allowable queue length \\
$M$ & Number of control messages \\
$m_A$ & Midpoint of the arrival-rate-based utility function \\
$m_B$ & Midpoint of the queue-based utility function \\
$\mu$ & Service rate (UEs/sec) \\
$N$ & Number of UEs \\
$P(t)$ & Penalty function \\
$\mathrm{P}$ & Resilience metric \\
$\rho$ & System utilization \\
$t$ & Time variable \\
$t_0$ & Initial time step of a signaling event \\
$t_d$ & Final time step a signaling event \\
$t_r$ & Time step of full recovery time after a disruption \\
$t_{\text{rec}}$ & Recovery time after signaling event\\
$t_{p,i}$ & Processing time for the $i$-th control plane message \\
$T_{\text{N-th, Monolithic}}$ & Total attachment delay for the $N$-th UE in monolithic RAN \\
$T_{\text{N-th, OpenRAN}}$ & Total attachment delay for the $N$-th UE in Open RAN \\
$u(t)$ & Utility function \\
$u_{\text{des}}(t)$ & Desired utility function \\
$V$ & Lyapunov parameter for utility maximization \\
$W$ & Lyapunov parameter for penalty \\
$w$ & Weighting factor \\
$\Delta L(t)$ & Lyapunov drift function \\
$\Delta t_{\text{des}}$ & Desired recovery time threshold \\
\hline
\end{tabular}%
}
\end{table}

\section{Related Works}\label{sec:rel_work}
This section reviews related works on RAN disaggregation (\ref{ssec:dis_lit}), signaling storm attacks (\ref{ssec:ssa_lit}), and network resilience (\ref{ssec:net_res_lit}), highlighting key challenges and open issues in Open RAN.

\subsection{Open RAN disaggregation}\label{ssec:dis_lit}
RAN disaggregation can be considered a pivotal concept in modern cellular networks, enabling flexible, vendor-agnostic systems through the decoupling of hardware and software. 
Ahmed et al.~\cite{ahmed2023} emphasize the potential of Open RAN in supporting autonomous systems and enhancing scalability, cost efficiency and interoperability through disaggregation and the introduction of functional splits. 
Wernet et al.~\cite{Wernet2023FaultTolerance} discussed the benefits of disaggregation in terms of fault-tolerance with the introduction of redundant RAN functions, to reduce downtime and improve service availability. 
Similarly, Bhattacharyya et al.~\cite{Bhattacharyya2023DisaggregatedRAN} showed how with the use of virtualized RAN components we can maintain service continuity and scalability. 
Hojeij et al.~\cite{Hojeij2024FlexiblePlacement} focuses on the flexible placement of O-CU and O-DU functionalities.
In consideration of the additional delays introduced by the use of open interfaces, Municio et al.~\cite{oran_tsn} propose Time-Sensitive Networking (TSN) solutions to meet stringent delay requirements for O-RAN interfaces.
Alavirad et al.\cite{alavirad2023} examined the O-RAN architecture with a focus on AI/ML-driven access control optimization, emphasizing the O-FH and F1 interfaces' role in ensuring interoperability and balancing functional splits for latency and performance. 
Finally, Baguer et al.~\cite{baguer2024attacking} analyze the expanded attack surface of disaggregated O-RAN architectures, investigating various security challenges. Among their findings, they demonstrate how delays or packet loss in the F1-C interface can severely disrupt user attachment, leading to failures and service degradation.
\textit{However, most research primarily focuses on optimization strategies, with only a small portion examining the impact of additional delays introduced by open interfaces and the security vulnerabilities of disaggregated architectures. This imbalance underscores the need for further research to comprehensively address these critical challenges. In our work, we model the delays introduced by the F1 interface and compare the performance of monolithic and Open RAN architectures.}

\subsection{Signaling Storm Attacks}\label{ssec:ssa_lit}
Signaling storm attacks have been studied across mobile network generations. These attacks exploit signaling (control plane) protocol vulnerabilities to overload network resources, leading to service disruptions and degraded network performance.
Early research in this domain focused mainly on traditional RANs and network architectures, such as LTE and 5G.
For example, Gelenbe et al.~\cite{ss_mobile_networks} proposed a timeout-based method to detect and mitigate signaling storm attacks in 5G mobile networks, supported by mathematical modeling and simulations showing reduced network overload. 
Similarly, Pavloski~\cite{pavloski2019detecting} proposed a collaborative approach between the cloud and the RAN to detect signaling storms, using both counter-based detection and bandwidth monitoring techniques.
Furthermore, Ettiane et al.~\cite{ettiane_controlplane} studied the vulnerabilities introduced by the RRC protocol. They highlight attack scenarios such as state transition manipulation and fake system information requests. 
Finally, Hu et al.~\cite{signalingsecurity_5GNAS} highlighted vulnerabilities in 5G Non-Access Stratum (NAS) signaling that can amplify signaling traffic, such as deregistration attacks, emphasizing the need to study core network signaling's role in exacerbating signaling storm attacks. 

Although these studies offer valuable information on the detection of signaling threats in traditional 5G architectures, more research is needed to understand signaling storm attacks in Open RAN. Its disaggregated architecture presents new challenges, increasing susceptibility to signaling overloads under high traffic, highlighting the need for tailored solutions to leverage Open RAN's benefits.
For example, Tabiban et al.~\cite{tabiban2023signaling} explored signaling storms in Open RAN, discussing threat models, existing mitigation strategies, architectural opportunities for improved detection and mitigation, and challenges due to disaggregation. 
Hoffmann and Kryszkiewicz~\cite{hoffmann2023signaling} proposed an xApp that monitors UE RRC messages and analyzes the timing advance (TA) parameter to detect abnormal activity during the initial device registration process.
Lastly, Mayhoub et al.~\cite{Mayhoub2024} introduced a new sub-use case for signaling storm attacks in Open RAN, exploiting compromised O-RUs to trigger excessive handovers and re-registrations, and subsequently used an ML-based detection system in the Near-RT RIC  to detect those attacks. 
\textit{While these studies have advanced the detection and mitigation of signaling storms in Open RAN, further research is needed to assess how its disaggregated architecture and diverse open-interface protocols increase susceptibility to such attacks. In this research we analyze how Open RAN architecture behave during mass UE attachment scenario due to the additional delays introduced by disaggregation. We also propose an adaptive mechanism to improve resilience of the system undergoing a signaling storm.}

\begin{table}[!t]
\centering
\caption{Comparison of Open RAN Signaling Storm Related Approaches}
\label{tab:signaling_storm_comparison}
\resizebox{\columnwidth}{!}{%
\begin{tabular}{|l|c|c|c|c|}
\hline
\textbf{Feature} & \textbf{\cite{tabiban2023signaling}} & \textbf{\cite{hoffmann2023signaling}} & \textbf{\cite{Mayhoub2024}} & \textbf{This Work} \\
\hline \hline
Open RAN Signaling Storm Attack Models & \checkmark & \checkmark & \checkmark & \checkmark \\
Analysis of Disaggregation Impact & \checkmark & \xmark & \xmark & \checkmark \\
Detection Mechanism Proposed &  \checkmark & \checkmark  & \checkmark & \xmark \\
Resilience Mechanism Proposed & \xmark & \xmark & \xmark & \checkmark \\
\hline
\end{tabular}%
}
\end{table}

\subsection{Network Resilience}\label{ssec:net_res_lit}
Resilience has become a pivotal concept across several disciplines, such as cyber-physical systems (CPS) and networking, encapsulating a system's ability to withstand, adapt, and recover from disruptions. 
Najarian and Lim~\cite{Najarian2019ResilienceMetrics} highlighted the critical role of resilience metrics in objectively assessing system performance. 
Also, Cassottana et al.~\cite{Cassottana2023CPSResilience}, provided an extensive review of resilience frameworks for CPSs, emphasizing the need for structured methodologies to assess the resilience of a system before and after a disruption. 
These principles are increasingly applied to cellular networks to ensure reliability and robustness. 
For instance, Kaada et al.~\cite{Kaada2022Resilience5G} proposed a resilience quantification framework for 5G RAN for coverage prediction and outage mitigation. 
In addition, Li et al.~\cite{Li2023ShortTerm} presented a model to dynamically evaluate network resilience under traffic changes, highlighting the importance of short-term resilience analysis. 
Furthermore, as networks transition from 5G to future generations, the adoption of resilience-by-design principles becomes essential. Khaloopour et al.~\cite{Khaloopour2024Resilience6G} introduced this concept, emphasizing the integration of adaptive, self-aware, and protective mechanisms across all network layers to proactively address emerging challenges.
Similarly, Reifert et al.~\cite{reifert2024resiliencecriticality} highlighted the importance of criticality-aware resilience in 6G networks, proposing strategies to prioritize essential services and develop tailored mechanisms, that adhere to certain restrictions, to ensure system functionality even under adverse conditions.
Finally, Han et al.~\cite{Han2017Resiliency} explore how virtual network functions (VNFs) can enhance the scalability and resilience of 5G systems due to their dynamic reconfigurability.  
\textit{Despite these significant contributions, there are still no official guidelines from standardization bodies on quantifying resilience. In this paper, we propose our utility function that can be used for queuing systems to compute the resilience of an Open RAN system in UE attachment scenarios.}

\begin{figure*}[!t]
    \centering
    \includegraphics[width=0.8\textwidth]{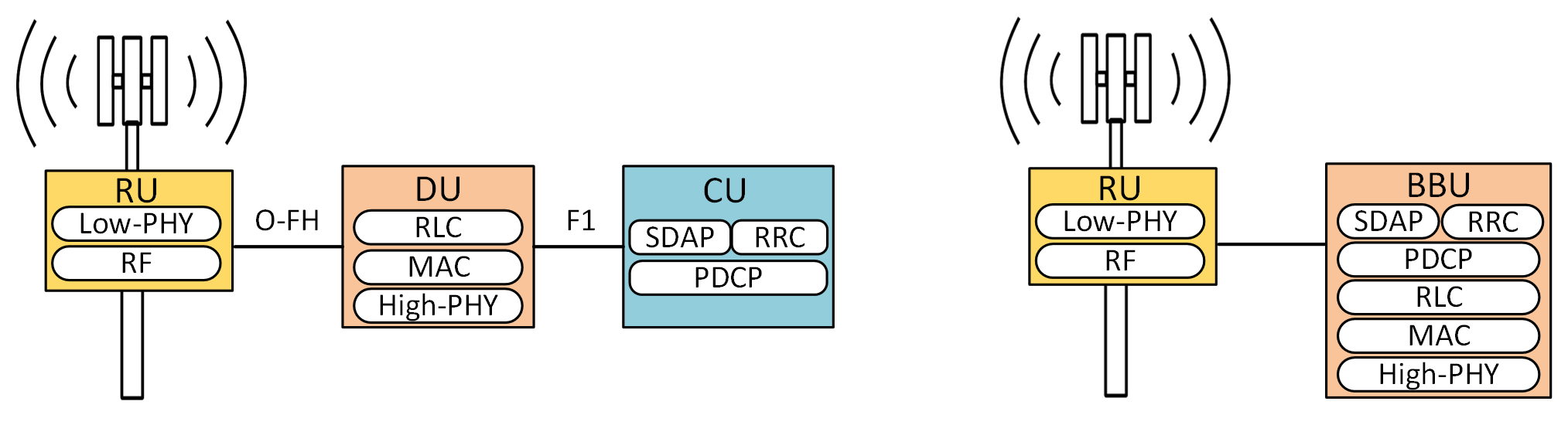} %
    \caption{Open RAN versus monolithic RAN architectures.}
    \label{fig:trad_vs_oran}
\end{figure*}

\begin{figure}[!t]
    \centering
    \includegraphics[width=0.8\columnwidth]{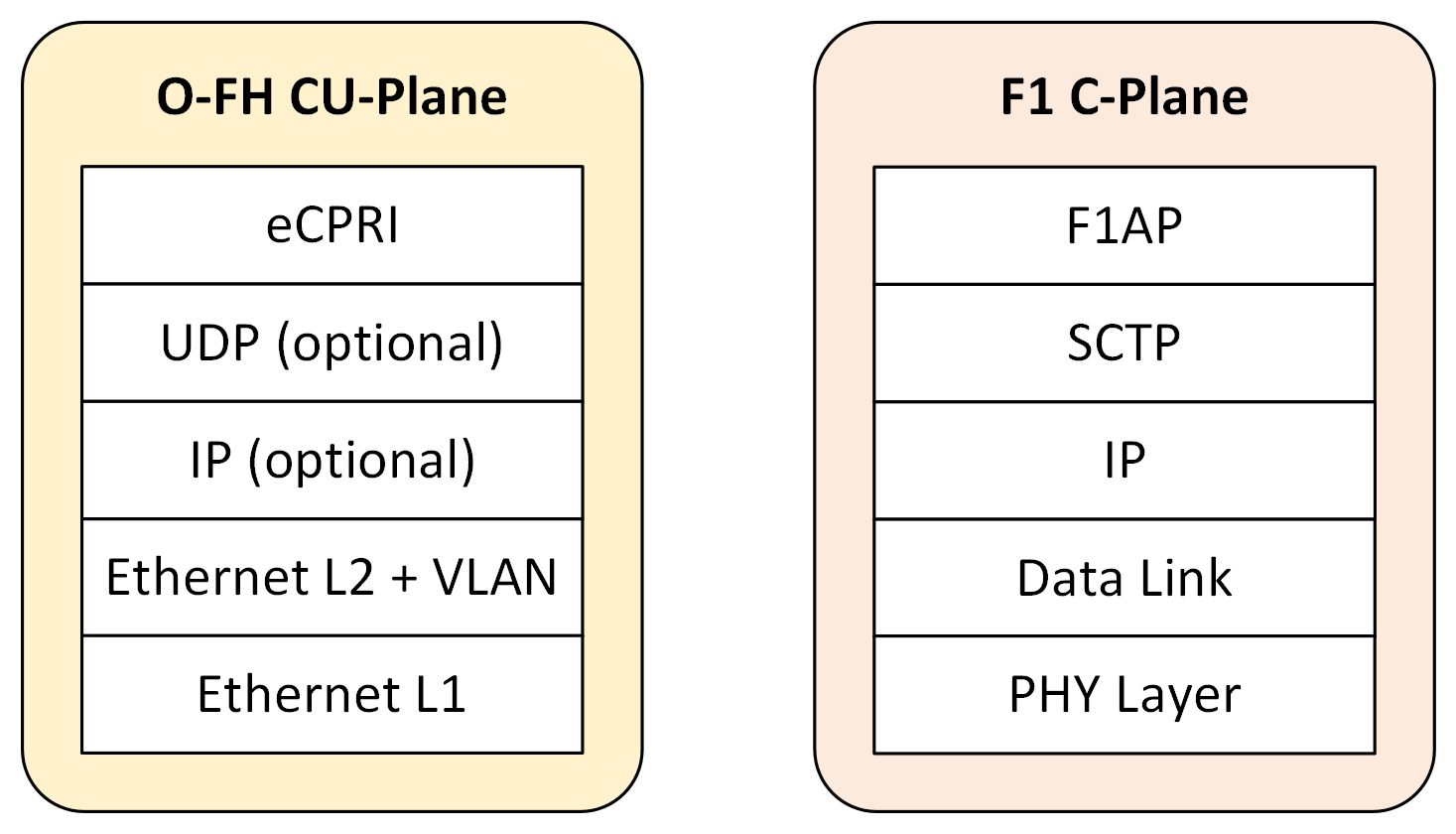} %
    \caption{O-FH and F1-C protocol stacks.}
    \label{fig:protocol_stack}
\end{figure}

\section{Open Radio Access Network}\label{sec:openran}

This section provides an overview of Open RAN architecture and key protocols. Section~\ref{ssec:acp} outlines its components, functional splits, and open interfaces, including the O-FH and F1. Section~\ref{ssec:rrc} details the UE initial attachment procedure, emphasizing key signaling processes and protocol overhead.

\subsection{Architecture, Components and Protocols}\label{ssec:acp}
Open RAN signifies the evolution from traditional RAN architectures towards an open, interoperable, disaggregated, and programmable framework. This transition introduces additional components and interfaces that play a fundamental role in Open RAN's functionality, as outlined in~\cite{ORAN_WG1_ARCH_DESC}. A pivotal aspect of this evolution is the disaggregation of the previously unified, vendor-specific base stations into distinct, modular components, aligning with 3GPP's proposals for the next-generation NodeBs (gNBs) in NR (New Radio)~\cite{3GPP38401_gNB}. The gNB is now separated into the RU, DU and CU. Disaggregating the base-band unit (BBU) may increase flexibility but also raise questions about function distribution across components. To address this, 3GPP has proposed several functional splits~\cite{3GPPTR38801_splits}, among which the 7.2x functional split adopted by the O-RAN alliance. Fig.~\ref{fig:trad_vs_oran} details the functions supported by each component in the O-RAN alliance. The RU handles physical layer (low PHY) processing, including RF signal processing and analog-to-digital conversion. The DU manages the rest of the physical layer (high PHY), along with the Medium Access Control (MAC) and Radio Link Control (RLC) layers, which are closely synchronized to facilitate data transfer. Finally, the CU oversees higher layers of the 3GPP stack: the RRC layer manages connection lifecycles, the Service Data Adaptation Protocol (SDAP) layer ensures Quality of Service for traffic flows, and the Packet Data Convergence Protocol (PDCP) layer is responsible for packet reordering, duplication, and encryption~\cite{Polese_O_RAN_arch}.

The ordinary function of the Open RAN is complemented by the use of open interfaces, that connect the various components of the Open RAN architecture, proposed both by the O-RAN Alliance and the 3GPP. The Open RAN interfaces can provide access to data analytics, enabling improved control, automation, and optimization within the RAN. The two interfaces responsible for the communication between the disaggregated RAN components are the open fronthaul (O-FH) interface that connects the RU and the DU, and the F1 interface that connects the DU with the CU. 

\begin{table*}[!t]
\centering
\caption{F1AP Information Elements based on ETSI TS 138 473 V15.8.0.}
\label{tab:f1ap-ies}
\begin{tabular}{|c|c|c|}
\hline
\textbf{RRC Message} & \textbf{Information Element and Description} & \textbf{Overhead} \\ \hline\hline
\makecell[l]{Initial UL RRC Message \\ RRC Setup Request} 
    & \makecell[l]{\textbf{Message Type}: Uniquely identifies the message being sent, mandatory for all messages. \\ 
                  \textit{Procedure Code}: Mandatory, Integer (0..255), represents the specific procedure. \\ 
                  \textit{Type of Message}: Mandatory, Choice of types (e.g., Initiating Message, Successful Outcome, etc.). \\
                  \textbf{gNB-DU UE F1AP ID}: Uniquely identifies the UE association within the gNB-DU, Integer (0..$2^{32}$-1). 
                  \\ \textbf{NR-CGI}: Used to globally identify a cell. \\
                      \textit{PLMN Identity}: OCTET STRING (SIZE(3)) \\
                      \textit{NR Cell Identity}: BIT STRING (SIZE(36)) \\
                  \textbf{C-RNTI}: Integer (0..65535), uniquely identifies the UE, as defined in TS 38.331. \\
                  \textbf{RRC Container}: Contains the RRCSetupRequest message payload. \\ 
                  \textbf{Transaction ID}: Uniquely identifies a procedure among parallel procedures of the same type, Integer (0..255).} 
    & 16 Bytes \\ \hline

\makecell[l]{DL RRC Message \\ RRC Setup} 
    & \makecell[l]{\textbf{Message Type} \\ 
                  \textbf{gNB-CU UE F1AP ID}: Uniquely identifies the UE association within the gNB-CU, Integer (0..$2^{32}$-1). \\ 
                  \textbf{gNB-DU UE F1AP ID} \\ 
                  \textbf{SRB ID}: Identifier for the Signaling Radio Bearer of a UE, Integer (0..$3$) \\ 
                  \textbf{RRC Container}: Contains the RRCSetup message payload.} 
    & 10 Bytes \\ \hline

\makecell[l]{UL RRC Message \\ RRC Setup Complete} 
    & \makecell[l]{\textbf{Message Type} \\ 
                  \textbf{gNB-CU UE F1AP ID}\\ 
                  \textbf{gNB-DU UE F1AP ID} \\ 
                  \textbf{SRB ID} \\ 
                  \textbf{RRC Container}: Contains the RRCSetupComplete message payload.}
    & 10 Bytes \\ \hline
\end{tabular}
\end{table*}

The O-FH interface consists of four planes: control (C), user (U), synchronization (S), and management (M), each designed to provide specific functionalities. The C-plane is responsible for transferring commands from the high-PHY layer of the DU to the low-PHY layer of the RU, including scheduling and beam-forming settings. The U-plane's primary role is to transmit I/Q samples, which are the modulated transmitted information, in the frequency domain between the RU and DU. The S-plane ensures synchronization of time, frequency, and phase between the DU and RU clocks, maintaining the proper functioning of a distributed time and frequency-slotted system. Lastly, the M-plane manages the initialization and management of the connection between the DU and RU, as well as the configuration of the RU~\cite{Polese_O_RAN_arch}. Each plane of the O-FH interface uses different protocol stacks as well as different security protocols. This study focuses on the C/U planes since are the ones used to carry the signals from the RU to the DU. The O-FH C/U planes use the protocol stack illustrated in Fig.~\ref{fig:protocol_stack}. At the top of the stack, the eCPRI (enhanced Common Public Radio Interface) protocol is used to transport user-plane and control-plane data in a packet-based manner, adding 4 bytes of overhead~\cite{ecpri2019}. Below eCPRI, optional UDP (User Datagram Protocol) and IP (Internet Protocol) layers may be included. Ethernet Layer 2 (L2) provides framing and addressing with MAC addresses and may incorporate VLAN tagging. Finally, Ethernet Layer 1 (L1) manages the transmission of raw bitstreams over the physical medium, typically through optical fiber cables~\cite{cho2021secure}. Due to strict latency requirements the O-RAN alliance has not mandated the use of any security protocols to be used in the O-FH interface. However, researchers are currently examining the use of MacSec as a possible solution~\cite{O_FH_macsec}.

\begin{figure}[!t]
    \centering
    \includegraphics[width=0.5\columnwidth]{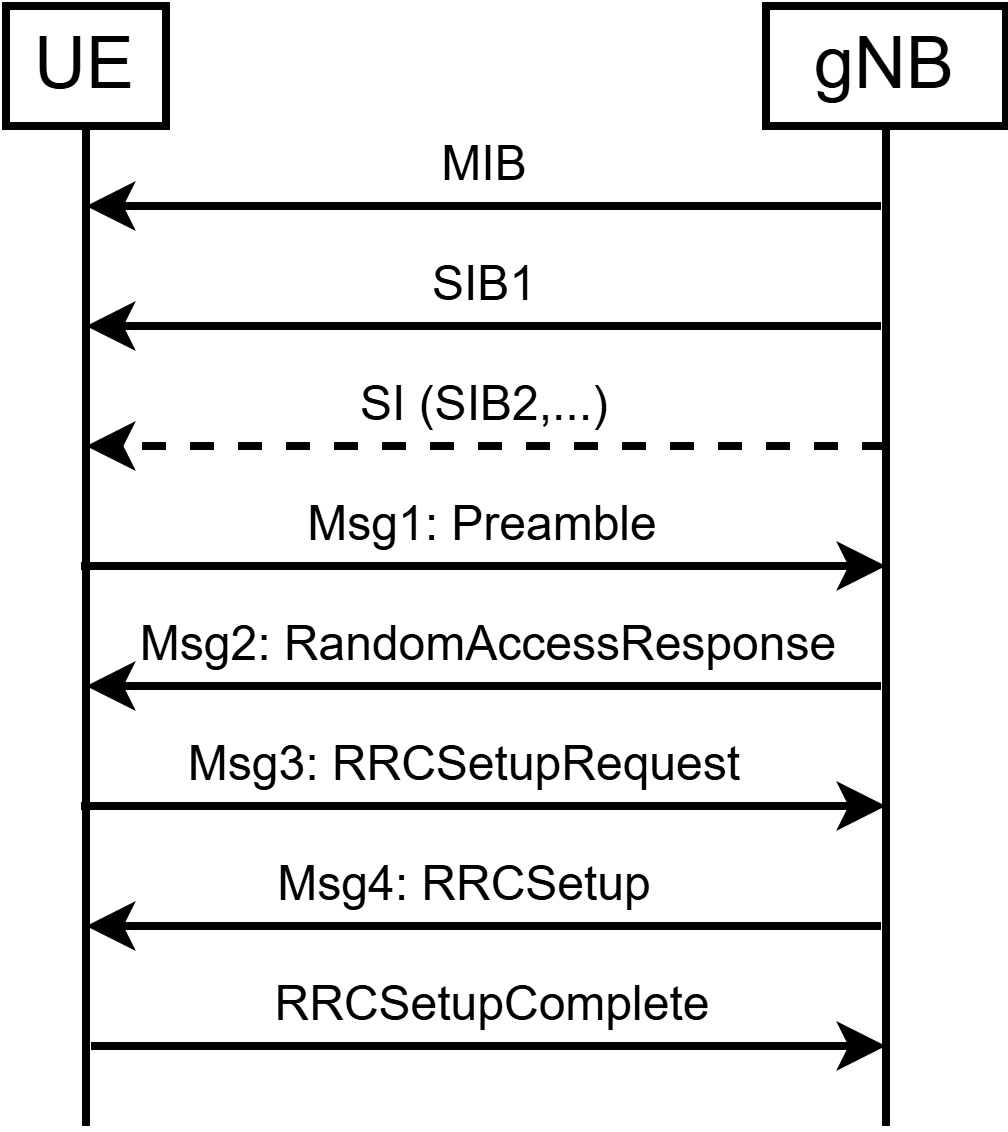} 
    \caption{5G NR Initial UE attachment procedure.}
    \label{fig:initial_attachment}
\end{figure}

Furthermore, the F1 interface, which connects the DU and CU, is standardized by 3GPP~\cite{3gpp_ts_38_470}. It is divided into two planes: F1-C (control plane), used for signaling messages between the RAN components, and F1-U (user plane), which handles user data transfer. In this paper, we focus on the F1-C interface which is responsible for transferring the RRC signal messages, sent from the UE to the CU (via the DU), for processing. The F1-C protocol stack, as shown in Fig.~\ref{fig:protocol_stack}, consists of the F1 Application Protocol (F1AP) at the top layer, which manages signaling exchanges. The F1AP protocol introduces additional overhead to the transmitted data between the CU and the DU. There are three distinct types of overhead, each associated with a different type of RRC message~\cite{etsi_ts_138_473_v1580}. The specific overhead for each message and its information elements are detailed in Table~\ref{tab:f1ap-ies}.

Additionally, these messages are encapsulated into a PDCP packet data unit (PDU), which introduces around 8 bytes of overhead. Next, the SCTP (Stream Control Transmission Protocol) is used for reliable message transport. SCTP adds at least 28 bytes of overhead—12 bytes for the common header and 16 bytes for the data chunk header. Additionally, SCTP applies padding to ensure the payload size is a multiple of 4 bytes. Following this, the IP layer handles routing, contributing 20 bytes of overhead for IPv4 or 40 bytes for IPv6. Finally, the Data Link layer (with 18 bytes of overhead for Ethernet) and the Physical layer (with 9 bytes of overhead) provide the necessary framing and physical transmission across the network.

The F1-C interface, as specified by 3GPP, should support confidentiality, integrity, and replay protection~\cite{3gpp_ts_33_501}. This can be achieved through mechanisms such as IPsec (Internet Protocol Security) or TLS (Transport Layer Security). Both protocols introduce additional overheads through encryption and encapsulation (25-40 bytes for TLS, 57 bytes for IPSec), which result in delays in the transmission of the payload~\cite{groen2024securing}.

\begin{table*}[!t]
\centering
\caption{RRC Messages Information Elements based on 3GPP TS 38.331.}
\label{tab:rrc_ies}
\begin{tabular}{|p{0.22\textwidth}|c|p{0.22\textwidth}|c|p{0.22\textwidth}|c|}
\hline
\multicolumn{2}{|c|}{\textbf{RRC Setup Request}} & \multicolumn{2}{c|}{\textbf{RRC Setup}} & \multicolumn{2}{c|}{\textbf{RRC Setup Complete}}  \\ \hline\hline
\multicolumn{1}{|c|}{\textbf{IE}} & \textbf{Size} & \multicolumn{1}{|c|}{\textbf{IE}} & \textbf{Size} & \multicolumn{1}{|c|}{\textbf{IE}} & \textbf{Size} \\ \hline
\multicolumn{1}{|l|}{\begin{tabular}[c]{@{}l@{}}UE-identity \\ \hspace{1em}ng-5G-S-TMSI-Part1\\ \hspace{1em}random value\end{tabular}} & 39 bits & RRC\_TransactionIdentifier & 2 bits & RRC\_TransactionIdentifier & 2 bits \\ \hline
\multicolumn{1}{|l|}{\begin{tabular}[c]{@{}l@{}}Establishment\_Cause \\ \hspace{1em}emergency\\ \hspace{1em}highPriorityAccess\\ \hspace{1em}...\end{tabular}} & 4 bits  & \begin{tabular}[c]{@{}l@{}}radioBearerConfig\{\\ \hspace{1em}srb-ToAddModList,\\ \hspace{1em}drb-ToAddModList,\\ \hspace{1em}securityConfig\}\end{tabular} & 9-13 bits & selectedPLMN-Identity & 3 bits \\ \hline
Spare & 1 bit & \begin{tabular}[c]{@{}l@{}}masterCellGroup\{\\ \hspace{1em}CellGroupConfig\}\end{tabular} & 128 bits & dedicatedNAS-Message & 80-200 bits \\ \hline
\end{tabular}
\end{table*}

\subsection{UE Initial Attachment Procedure}\label{ssec:rrc}
The initial attachment of a UE to the gNB (5G base station) involves the exchange of several key messages, as illustrated in Fig.~\ref{fig:initial_attachment}. The procedure begins with the gNB broadcasting the Master Information Block (MIB). The MIB contains critical information such as the system frame number, common sub-carrier spacing, Synchronization Signal Block (SSB) sub-carrier offset, and cell barring status. This information is essential for downlink synchronization and helps the UE locate and decode subsequent messages. Following the MIB, the gNB broadcasts the System Information Block 1 (SIB1). SIB1 provides the UE with the necessary details for accessing the cell. It includes scheduling information, cell identification details such as the Public Land Mobile Network (PLMN) and Tracking Area Code (TAC), and Random Access Channel (RACH) configuration. The MIB and SIB1 ensure the UE can synchronize, access, and communicate with the gNB. If additional SIBs are needed, they are broadcast by the gNB according to the network's configuration.

The second stage of the attachment procedure is the RACH process, which is responsible for establishing uplink synchronization. The UE initiates communication by sending message 1 (Msg1) to the gNB, containing the RACH preamble, a random access preamble from predefined sequences (i.e., Zadoff-Chu sequences). This preamble is referenced with the Random Access Preamble Id (RAPID) and transmitted using the physical RACH (PRACH). Upon receiving the preamble, the gNB sends message 2 (Msg2), called the Random Access Response (RAR). The RAR contains valuable information such as the timing advance (TA) for timing adjustment, the Random Access-Radio Network Temporary Identifier (RA-RNTI) for the UE, and an initial uplink grant. Messages 1 and 2 involve only the RU and DU of the gNB, that contain the PHY layer, following the same procedure as in monolithic RANs. 

Following the uplink allocation from Msg2, the UE transmits message 3 (Msg3), the RRC Setup Request. From this point onward, the remainder of the UE initial attachment messages are handled at the CU, responsible for managing the RRC procedure. The RRC Setup Request message includes several elements to initiate the RRC connection. First, the \textit{UE-identity}, which can either be the ng-5G-S-TMSI-Part1 (a 39-bit string used if the UE is already known to the network) or a random value (a 39-bit string used when the UE is initiating access for the first time). The \textit{Establishment\_Cause} field (4 bits) indicates the reason for the RRC connection request, such as mobile-originated signaling, data, or voice call. Finally, there is a \textit{spare} 1-bit field reserved. The RRC Setup Request message is 44 bits long (6 bytes).

The gNB then sends message 4 (Msg4), the RRC Setup, which configures the radio connection between the UE and gNB. It includes the \textit{RRC\_TransactionIdentifier} (2 bits) to identify the transaction, and the RRC Setup information elements, which consist of the \textit{radioBearerConfig} (9–13 bits) for configuring the signaling radio bearers (SRBs) and data radio bearers (DRBs), and the \textit{masterCellGroup} (128 bits) containing the CellGroupConfig. Optional fields can add up to 24 bits. The total size of the message is around 18 bytes without optional fields and up to 22 bytes with all optional fields included.

Finally, the UE sends the RRC Setup Complete message to the gNB, completing the RRC connection establishment and starting the connection with the core. This message includes fields like the \textit{RRC\_TransactionIdentifier} and the \textit{selectedPLMN\_Identity} (3 bits). The message also carries the \textit{dedicatedNAS\_Message}, which varies in size, typically ranging from 80 to 200 bits depending on the specific NAS message type. Optional fields can also be included. The total size of the message is approximately 23 bytes without optional fields, and up to 32 bytes if all optional fields are included. Table~\ref{tab:rrc_ies} details the information elements included in each message~\cite{3gpp-ts-38.331,5gnr_ns3, eventhelix_5g_registration}.

\section{UE Initial Attachment Procedure Analysis}\label{sec:ueia}
This section examines the factors affecting UE initial attachment in Open RAN. Section~\ref{ssec:md} models key delay components, while Section~\ref{ssec:da_pe} analyzes performance under varying load conditions. Section~\ref{ssec:massue} examines mass UE attachment scenarios and their impact on network congestion. Finally, Section~\ref{ssec:res} explores network resilience and adaptation strategies to mitigate signaling storms.

\subsection{Modeling of Delays}\label{ssec:md}
The UE initial attachment procedure involves multiple signaling exchanges between the network entities to establish a connection. In traditional monolithic RAN, these exchanges occur within a tightly integrated system, minimizing additional transmission delays. However, in Open RAN architectures, the introduction of the open interfaces between the RUs, DUs and CUs can introduce extra latency, affecting the overall attachment time. This paper assumes that the O-FH protocol stack is identical for both monolithic RAN and Open RAN, with a maximum allowed one-way delay of 250 \(\mu\)s~\cite{3gpp-tr-38.801}, with any additional delays attributed to introducing the F1 interface. To quantify these delays, we analyze the impact of control plane signaling, associated protocol overheads, and potential increases in processing and transmission times across the CU-DU split. The delays considered are:

\begin{itemize}    
    \item \textbf{Transmission Delay ($D_{t}$):} The time required to send a message over the F1-C interface, accounting for protocol overhead, given by \( D_{t} = \frac{S + O}{R} \). Here, the message size (\( S \)) includes RRC signaling messages, the protocol overhead (\( O \)) represents additional data added by communication protocols, and the transmission rate (\( R \)) denotes the capacity of the F1-C link, typically measured in megabits per second (Mbps). Table~\ref{tab:rrc_message_sizes} lists the sizes of RRC messages exchanged (including other protocol headers) during the initial UE attachment procedure.

    \begin{table}[!t]
    \centering
    \caption{RRC Message Sizes (in Bytes) During Initial UE Attachment with TLS and IPSec Overheads.}
    \label{tab:rrc_message_sizes}
    \begin{tabular}{|c|c|c|}
    \hline
    \textbf{RRC Message} & \textbf{Size with TLS} & \textbf{Size with IPSec} \\
    \hline \hline
    RRC Setup Request & 124–139 & 156 \\
    RRC Setup & 128–143 & 160-164 \\
    RRC Setup Complete & 136–151 & 168-174 \\
    \hline
    \end{tabular}
    \end{table}
    
    \item \textbf{Propagation Delay ($D_{p}$):} The time taken for a signal to travel between the DU and CU is given by \( D_{p} = \frac{d}{v} \). The propagation speed (\( v \)) depends on the type of link used, and the distance (\( d \)) between the DU and CU can vary. For example, the propagation speed of fiber optics cable is approximately \( 2 \times 10^8 \, \text{m/s} \). The added latency per 100 meters is \( 0.5\,\mu s \).

    \item \textbf{Queuing Delay ($D_{q}$):} The time a message spends in queues before processing or forwarding. In normal scenarios where the arrival rate of UEs attempting to attach to the network is less than the attachment service rate, the queuing delay is almost negligible ($D_{q} \approx 0$).
    
    \item \textbf{Processing Delay ($D_{r}$):} The time taken by the DU and CU to process and forward messages, including security-related operations such as encryption/decryption and integrity checks, which are embedded within the $D_{\text{r\_CU}}$ and $D_{\text{r\_DU}}$ components
    \begin{equation}
        D_{r} = D_{\text{r\_DU}} + D_{\text{r\_CU}},
    \end{equation}
     The processing delay can be approximated as
    \begin{equation}
        D_{r} \approx D_{\text{total}} - (D_{t} + D_{p} + D_{q}).
    \end{equation}
\end{itemize}
 Moreover, the total allowed one-way delay between the DU and CU ranges between 1.5 ms and 10 ms ~\cite{3gpp-tr-38.801}. Given that the $D_{t}$ and $D_{p}$ are typically only a few microseconds, and assuming a typical scenario where the $D_{q}$ is nearly negligible ($D_{q} \approx 0$), we can extract that the $D_{r}$ occupies the majority of the total delay. In a scenario where the total delay reaches the lower bound of 1.5 ms, the processing delay $D_{r}$ can be approximated to be close to this lower bound.

The overall delay introduced by the F1-C interface during the initial UE attachment, considering security measures, is given by
\begin{equation}
    D_{\text{total}} = \sum_{i=1}^M (D_{\text{t,i}} + D_{p} + D_{\text{r,i}} + D_{\text{q,i}}),
    \label{eq:delays_sec}
\end{equation}
where $M$ is the total number of control plane messages exchanged during the attachment process. The delays introduced are linear in nature, as they scale directly with the size of each message, the number of messages exchanged, and the processing time required by the CU and DU. This linear relationship implies that, even with the introduction of disaggregation and the addition of security protocols, the total delay will increase proportionally with the number and size of control plane messages.

\subsection{Delay Accumulation and Performance Evaluation Under Varying Load Conditions}\label{ssec:da_pe}
Having established the fundamental delay components introduced by the F1 interface, we now examine how these delays accumulate as multiple UEs attempt to attach to the network. Unlike an isolated UE attachment, where delays remain relatively small, real-world deployments involve multiple UEs joining the network simultaneously, leading to potential queuing effects and increased attachment times. This subsection assumes that UEs attach sequentially, meaning each UE must complete its attachment process before the next one can begin. Consequently, in a scenario where $N$ UEs are queued for network attachment, the delay for each subsequent UE increases as more UEs join the queue. Under this assumption, we can compute the total attachment delay for the $N$-th UE in both monolithic RAN and Open RAN setups. We assume a one-way delay of 0.25 ms between the RU and the BBU for the monolithic RAN setup. This delay corresponds to Split Option 7, where the division occurs between the low and high physical layers of RU and BBU respectively. In the Open RAN setup, the total one-way delay from the RU to the Central Unit (CU) is assumed to be 1.75 ms. This consists of 0.25 ms from the RU to the DU, based on Split Option 7, and an additional 1.5 ms from the DU to the CU, corresponding to Split Option 2, which splits between the RLC and PDCP layers~\cite{3gpp-tr-38.801}. Thereby, we have
    \begin{equation}
        \label{eq:t_mono}
        T_{\text{N-th, Monolithic}} = N \times \sum_{i=1}^{M} \left( t_{p,i} + D_{\text{RU-BBU}} \right),
    \end{equation}
    
    \begin{equation}
        \label{eq:t_open}
        T_{\text{N-th, OpenRAN}} = N \times \sum_{i=1}^{M} \left( t_{p,i} + D_{\text{RU-CU}} \right),
    \end{equation}
where
\begin{itemize}
    \item \( N \): The number of UEs in the queue, representing the cumulative effect of sequential processing.
    \item \( M \): The total number of control plane messages exchanged during the attachment process.
    \item \( t_{p,i} \): The processing time for the \( i \)-th control plane message.
    \item \( D_{\text{RU-BBU}} \): The one-way delay between the RU and BBU for each message in monolithic RAN.
    \item \( D_{\text{RU-CU}} \): The one-way delay from RU to CU (via DU) for each message in Open RAN.
\end{itemize}
It is important to note that \( \sum_{i=1}^{M} t_{p,i} \) is identical for both monolithic RAN and Open RAN, as each control plane message requires the same internal processing time regardless of the architecture. Thus, the primary difference in the total delay experienced by the $N$-th UE between the monolithic RAN and Open RAN setups arises from the additional one-way delay specific to the Open RAN architecture.

In queuing systems, as the arrival rate (\( \lambda \)) approaches the service rate (\( \mu \)), system utilization increases, leading to longer queues and higher delays. To characterize these dynamics, we define the service rates for monolithic RAN and Open RAN as follows
\begin{equation}
    \mu_{\text{Monolithic}} = \frac{1}{\sum_{i=1}^{M}( t_{p,i} + D_{\text{RU-BBU}})},
    \label{eq:sr_mono}
\end{equation}

\begin{equation}
    \mu_{\text{OpenRAN}} = \frac{1}{\sum_{i=1}^{M}( t_{p,i} + D_{\text{RU-CU}})}.
    \label{eq:sr_open}
\end{equation}
The arrival rate \( \lambda \) varies from low values up to near the service rate in both setups. Using an M/M/1 queue model, the following key performance metrics were subsequently used to characterize the behavior of the system
\begin{itemize}
    \item \textbf{Utilization} (\( \rho \)): Represents the fraction of time the system is occupied, calculated as \( \rho = \frac{\lambda}{\mu} \).
    
    \item \textbf{Expected Number of UEs in the System} (\( L_s \)): This includes both UEs in the queue and those being processed, given by \( L_s = \frac{\rho}{1 - \rho} = \frac{\lambda}{\mu - \lambda} \).
    
    \item \textbf{Average Delay per UE} (\( W \)): The average time each UE spends in the system (including waiting and processing time) is \( W = \frac{1}{\mu - \lambda} \).
\end{itemize}

\subsection{Mass UE Attachment Scenario} \label{ssec:massue}
In high-density scenarios the network can experience significant congestion. Under normal utilization rates, the average time a UE spends in the system before completing its attachment remains within an acceptable range. However, as the network load increases, attachment delays grow significantly. When the system approaches saturation, each additional UE can further amplify delays due to queue buildup.

A signaling storm is a situation where the arrival rate exceeds the service rate, leading to congestion. In such cases, when \( \lambda > \mu \), queue lengths increase, causing prolonged attachment times. Monolithic RAN (\(\mu_{\text{Monolithic}} = 32.52\) UEs/s) and Open RAN (\(\mu_{\text{OpenRAN}} = 28.37\) UEs/s) operate under different service rate capacities, which influence their ability to handle mass UE attachment scenarios. An example of such a scenario is when the initial arrival rate of \( \lambda_{\text{normal}} = 20 \) UEs/s suddenly spikes to \( \lambda_{\text{storm}} = 200 \) UEs/s for a fixed duration before returning to normal levels. This process consists of three phases: ramp-up, steady-state, and ramp-down. The time required for the system to recover from the storm depends on how quickly the queues deplete once the arrival rate falls below the service rate.

In addition to sequential UE attachment, a more realistic approach considers parallel attachment, where multiple UEs perform the procedure simultaneously. This scenario can be modeled as an M/M/c queue, where \( c \) represents the number of servers handling attachment requests concurrently. The utilization in this case is given by \( \rho = \frac{\lambda}{c \mu} \). When the arrival rate exceeds the total system capacity (\( \lambda > c \mu \)), the system becomes unstable, leading to an indefinite increase in queue length (\( \rho > 1 \)). To describe the queue behavior over time during such an event, the queue length and waiting time can be expressed as
\( L_q(t+1) = L_q(t) + (\lambda(t) - c \mu) \),
and \( W_q(t+1) = \frac{L_q(t+1)}{c \mu} \).
These expressions describe how queue buildup and waiting time evolve in response to fluctuating arrival rates. The ability of a system to accommodate mass UE attachment depends on the number of servers available and the overall processing capacity, affecting how quickly queues are depleted following a congestion event.
 
\subsection{Resilience}\label{ssec:res}
Signaling storms, whether triggered by malicious activities—such as attacks from a UE botnet or a rogue base station—or by high-density user scenarios, pose a significant threat to the resilience of Open RAN networks. These events can lead to service degradation, network congestion, and even outages, challenging the network's ability to maintain its functionality. To address these risks effectively, Open RAN must embrace a comprehensive resilience framework. Principles include anticipation, absorption, adaptation during unforeseen conditions, and timely recovery after disastrous events.

The resilience behavior of the network can be quantitatively assessed using the $A^{3}\mathrm{RT}$ resilience metric, shown in (\ref{eq:res}), as proposed by \cite{reifert2024resiliencecriticality}. This metric evaluates resilience through three key components:
\begin{itemize}
    \item \textbf{Absorption:} Measuring the network's ability to resist and minimize the impact of disruptions during the time interval $[t_0, t_d]$.
    \item \textbf{Adaptation:} Reflecting the effectiveness of mechanisms that restore partial functionality in the interval $[t_d, t_r]$.
    \item \textbf{Time-to-Recovery:} Evaluating how quickly the network returns to a stable operational state. The recovery time $t_\text{rec}$ is defined as 1 if the recovery time is within the desired threshold ($\Delta t_\text{des}$), or is scaled proportionally otherwise.
\end{itemize}

\begin{equation}\label{eq:res}
\begin{aligned}
\mathrm{P} &= w_1 
\frac{\int_{t_0}^{t_d} u(t) \, dt}{\int_{t_0}^{t_d} u_{\text{des}}(t) \, dt}
+ w_2 
\frac{\int_{t_d}^{t_r} u(t) \, dt}{\int_{t_d}^{t_r} u_{\text{des}}(t) \, dt}
+ w_3 
\cdot t_\text{rec}, \\
t_\text{rec} &= 
\begin{cases} 
1, & t_r - t_0 \leq \Delta t_\text{des}, \\
\frac{\Delta t_\text{des}}{t_r - t_0}, & \text{otherwise}.
\end{cases}
\end{aligned}
\end{equation}
Here, \( w_1, w_2, \) and \( w_3 \) are weighting factors that balance the contribution of each resilience component, reflecting the importance of absorption, adaptation, and recovery in the overall resilience assessment. Their values should be determined based on service criticality, risk assessment, and system objectives, as suggested in \cite{Cassottana2023CPSResilience}. In high-risk scenarios, a larger \( w_1\) and \( w_2\) prioritizes the proactive measures of anticipation and absorption, while a higher \( w_3 \) emphasizes recovery when errors are rare or less harmful. A balanced configuration, such as \( w_1 = 0.4, w_2 = 0.4, w_3 = 0.2 \), ensures robust defense mechanisms while maintaining efficient recovery. The selection of these weights can be further refined based on empirical analysis and performance evaluation. Also, \( u(t) \) represents the utility function value over time, and \( u_{\text{des}}(t) \) denotes the desired utility level under ideal conditions.

To characterize the performance of Open RAN under varying loads, including signaling storms, and to describe the system's resilience, we propose a utility function. This function measures the overall stability and health of the system and is defined as follows
\begin{equation}
\begin{split}
    u(t) &= w_A u_A(t) + w_B u_B(t), \\
    u_A(t) &= \frac{1}{1 + \exp\bigl[k_A(\lambda(t) - m_A(t))\bigr]}, \\
    u_B(t) &= \frac{1}{1 + \exp\bigl[k_B(L_q(t) - m_B)\bigr]}. \\
\end{split}
\end{equation}
Where
\begin{itemize}
    \item $u_A(t)$ is the partial utility calculated based on the arrival rate $\lambda(t)$.
    \item $u_B(t)$ is the partial utility calculated based on the queue length $L_q(t)$.
    \item $w_A$ and $w_B$ are weighting factors, satisfying $w_A + w_B = 1$, that balance the contributions of $u_A(t)$ and $u_B(t)$. The values of these weights can be chosen based on empirical data or system requirements. For instance, if system stability is more critical, a higher $w_A$ (e.g., $w_A = 0.7$, $w_B = 0.3$) may be selected. Conversely, if controlling queue length is a higher priority, a higher $w_B$ (e.g., $w_A = 0.3$, $w_B = 0.7$) may be preferable. A balanced approach with $w_A = 0.5$ and $w_B = 0.5$ ensures equal consideration of both factors. 
    \item $\lambda(t)$ represents the arrival rate of requests at time $t$.
    \item $c(t)$ is the time-varying number of servers, and $\mu$ is the service rate per server.
    \item $m_A(t) = c(t) \mu \cdot m_{\text{frac}_A}$ denotes the midpoint of the arrival-rate-based utility function, dynamically adjusted based on $c(t)$, where $m_{\text{frac}_A}$ is a scaling factor that determines the fraction of $c(t) \mu$ used as the midpoint. In our case, we set $m_{\text{frac}_A} = 0.75$ to ensure that the utility function remains balanced, avoiding overly aggressive or conservative adjustments while maintaining stability in the system.
    \item $L_q(t) = \max\left( 0, L_q(t-1) + \lambda(t-1) - c(t-1) \mu \right)$ is the queue length of the system at time $t$.
    \item $L_{q\max}$ is the queue length threshold at which the partial utility function $u_B(t)$ drops to zero, indicating severe congestion.
   \item $m_B = L_{q\max} \cdot m_{\text{frac}_B}$ represents the midpoint of the queue-based utility function, where $m_{\text{frac}_B}$ is a scaling factor that determines the fraction of $L_{q\max}$ used as the midpoint. We set $m_{\text{frac}_B} = 0.5$ to achieve a more balanced utility value, ensuring that the function appropriately reflects the trade-off between queue length and utility value.
    \item $k_A > 0$ and $k_B > 0$ control the steepness of the transitions in $u_A(t)$ and $u_B(t)$, respectively.
\end{itemize}
The utility function $u(t)$ provides a value between 0 and 1, where higher values indicate greater system stability and resilience. By combining both arrival-rate-based and queue-length-based components,  the function captures the interplay between system load and congestion offering a holistic measure of resilience. 

To enhance the resilience of the network during signaling storms, we introduce an adaptation mechanism that dynamically adjusts the number of servers (\(c(t)\)) to stabilize the system while minimizing resource usage. This mechanism leverages Lyapunov's drift-plus-penalty method to dynamically determine the optimal \(c(t)\) at each time step, ensuring system stability and resilience while balancing resource efficiency and performance.

The Lyapunov's function quantifies system stability, and its drift is given by
\begin{equation} 
L(t) = \frac{1}{2} L_q(t)^2, \quad 
\Delta L(t) = L(t+1) - L(t).
\label{eq:lyapunov_drift}
\end{equation}
For \(L_q(t+1) > 0\), we approximate the drift as
\begin{equation} 
\Delta L(t) = L_q(t) \big[\lambda(t) - c(t)\mu\big] + \frac{1}{2} \big[\lambda(t) - c(t)\mu\big]^2.
\label{eq:drift_approx}
\end{equation}
To balance stability and performance, we introduce the penalty function
\begin{equation} 
P(t) = -V \cdot u(t) + W \cdot c(t),
\label{eq:penalty}
\end{equation}
where \(u(t)\) measures system utility, \(V\) prioritizes utility maximization, and \(W\) penalizes larger \(c(t)\). Thus, the drift-plus-penalty objective is
\begin{equation} 
\begin{split}
\Delta L(t) + P(t) &= L_q(t) \big[\lambda(t) - c(t)\mu\big] + \frac{1}{2} \big[\lambda(t) - c(t)\mu\big]^2 \\ &\quad - V \cdot u(t) + W \cdot c(t).
\end{split}
\label{eq:drift_penalty}
\end{equation}
The adaptation mechanism optimizes \(c(t)\) by solving
\begin{equation} 
\begin{split}
\min_{c(t)} \Big[ L_q(t) \big[\lambda(t) - c(t)\mu\big] &+ \frac{1}{2} \big[\lambda(t) - c(t)\mu\big]^2 \\ &- V \cdot u(t) + W \cdot c(t) \Big],
\end{split}
\label{eq:optimization}
\end{equation}
subject to
\[1 \leq c(t) \leq c_{\max}, \quad c(t) \in \mathbb{Z}^+.\]
By solving this optimization in real-time, the system dynamically adjusts \(c(t)\), ensuring it absorbs disruptions, adapts to changing conditions, and recovers efficiently.

\section{Quantitative Analysis and Discussion}\label{sec:qaad}

This section presents a quantitative assessment of delays and system behavior under varying load conditions in both monolithic RAN and Open RAN architectures. The analysis focuses on key performance metrics, including transmission delays, queuing behavior, and system utilization, to provide insights into how different network configurations respond to increasing traffic demands. Additionally, the impact of increasing network load on attachment delays is examined, along with mass UE attachment scenarios. Finally, network resilience and adaptive strategies to mitigate congestion and enhance system stability are evaluated.

\begin{figure}[!t]
\centering
\includegraphics[width=\columnwidth]{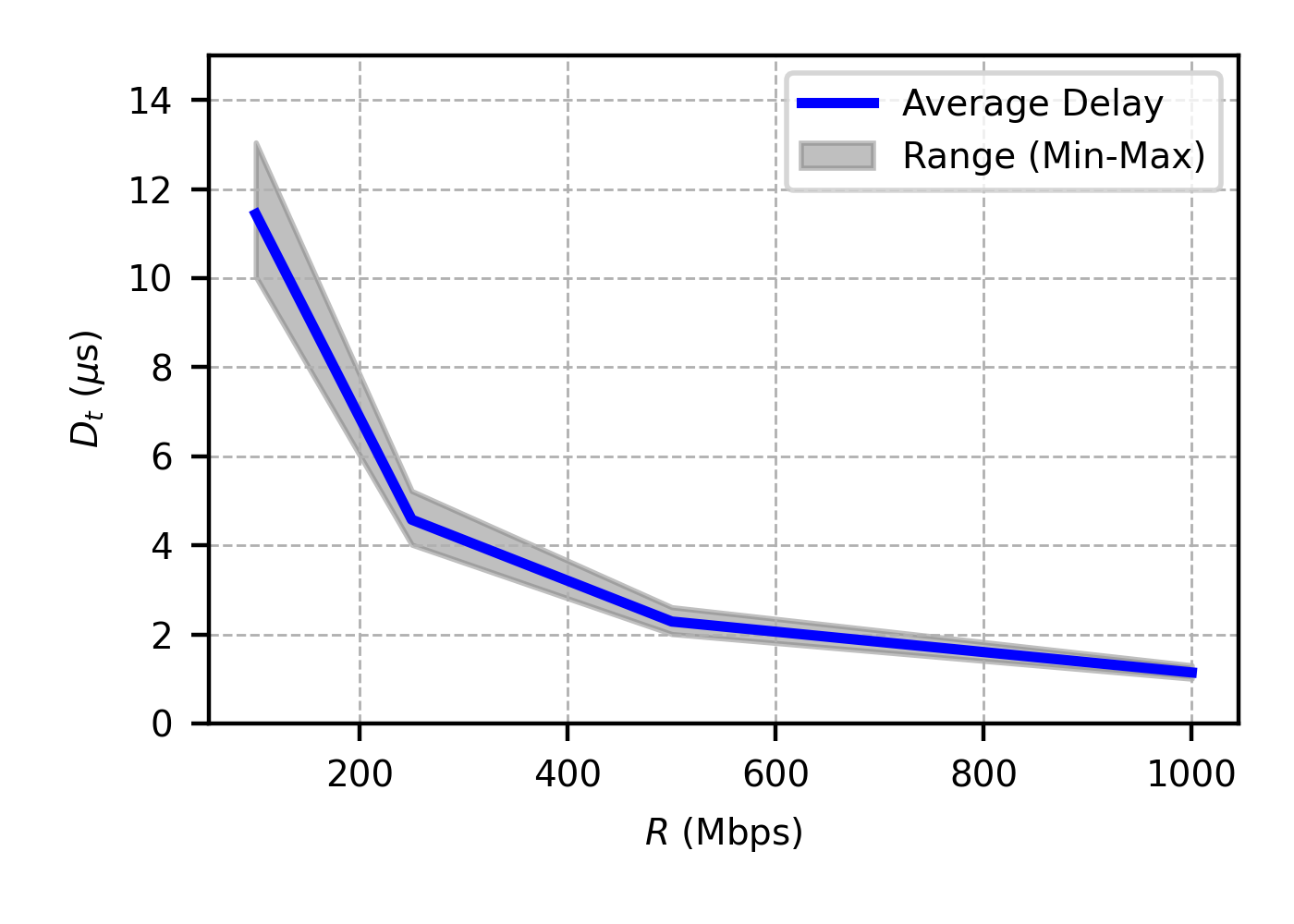}
\caption{Transmission Delay for RRC Messages with Various Transmission Rates.}
\label{fig:transmission_delay}
\end{figure}

We begin by examining the transmission delay across various transmission rates for all RRC messages and security protocols. Fig.~\ref{fig:transmission_delay} illustrates these variations, where the shaded area represents the range between the minimum and maximum delays, highlighting the variability caused by different message sizes and overheads.

\begin{table}[!t]
\centering
\caption{Performance metrics for monolithic and Open RAN configurations across varying load levels.}
\label{tab:perf_eval}
\begin{tabular}{|c|c|c|c|c|}
\hline
\( \mu \) (UEs/sec) & \( \rho \) & \( \lambda \) (UEs/sec) & \( L_s \) & \( W \) (ms) \\
\hline\hline
\multicolumn{5}{|c|}{\textbf{Monolithic RAN}} \\
\hline
32.52 & 0.1 & 3.25 & 0.11 & 34.16 \\
32.52 & 0.5 & 16.26 & 1 & 61.50 \\
32.52 & 0.9 & 29.27 & 9 & 307.7 \\
32.52 & 0.95 & 30.89 & 19 & 613.5 \\
\hline
\multicolumn{5}{|c|}{\textbf{Open RAN}} \\
\hline
28.37 & 0.1 & 2.84 & 0.11 & 39.17 \\
28.37 & 0.5 & 14.19 & 1 & 70.52 \\
28.37 & 0.9 & 25.53 & 9 & 352.5 \\
28.37 & 0.95 & 26.95 & 19 & 705 \\
\hline
\end{tabular}
\end{table}

\begin{figure}[!t]
\centering
\includegraphics[width=\columnwidth]{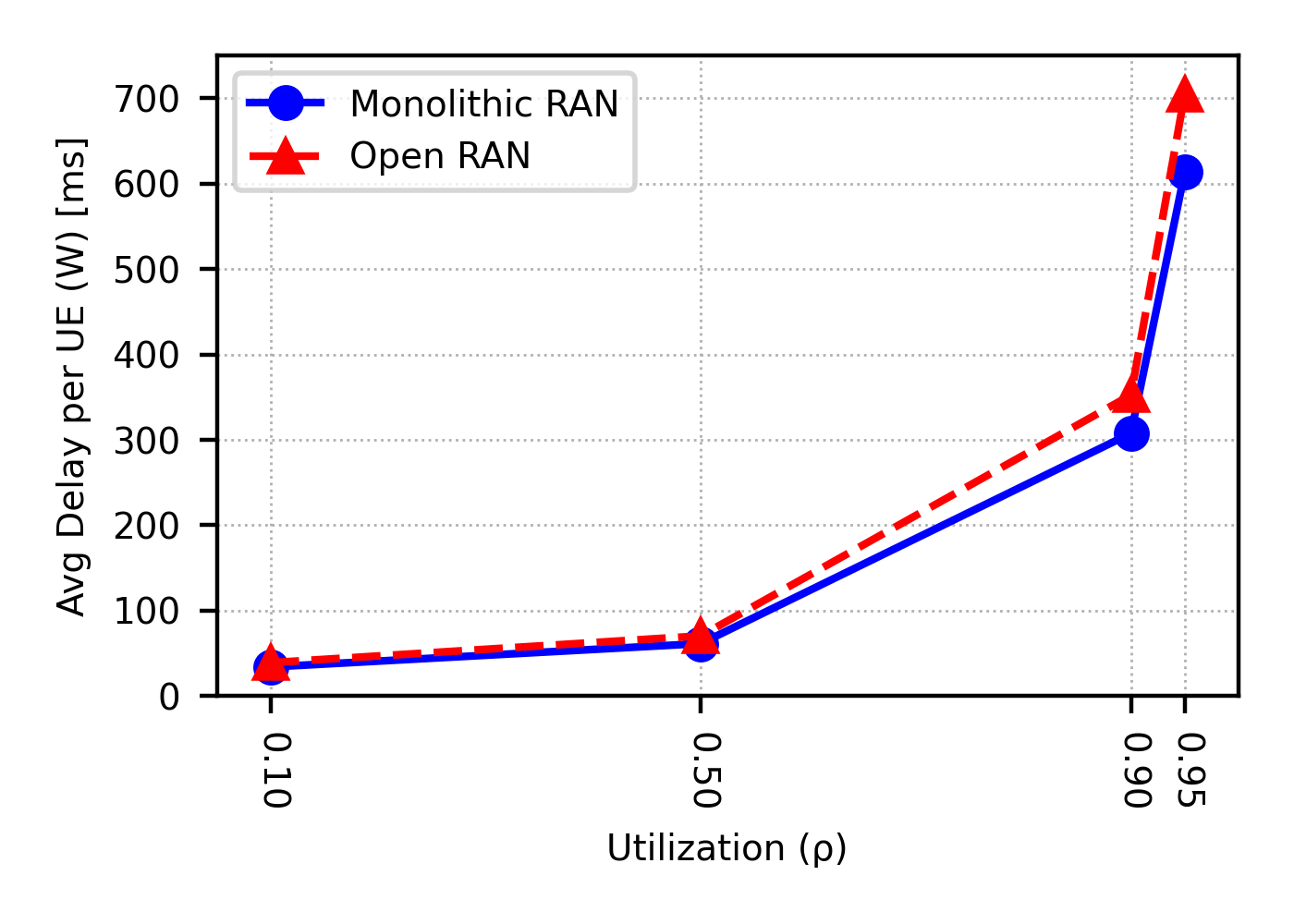}
\caption{Impact of Utilization (\( \rho \)) on Average Delay per UE (\( W \)).}
\label{fig:utilisation_vs_delay}
\end{figure}

Furthermore, to analyze the impact of increasing load on system behavior, we compare the performance of monolithic RAN and Open RAN configurations at different utilization levels. Table~\ref{tab:perf_eval} presents key performance metrics, while Fig.~\ref{fig:utilisation_vs_delay} visualizes the relationship between utilization (\(\rho\)) and average delay per UE (\(W\)). The results indicate that monolithic RAN consistently maintains lower overall delays under high-load conditions due to its higher service rate. However, both setups exhibit similar trends as load increases, with Open RAN experiencing slightly greater congestion as it approaches capacity.

In addition to evaluating system performance under varying loads, we analyze how the total internal message processing time (\( \sum_{i=1}^{M} t_{p,i} \)) influences key performance metrics, including utilization, queue length, and delay per UE. Service rates remain consistent with equations~(\ref{eq:sr_mono}) and~(\ref{eq:sr_open}). To isolate the effect of \( \sum_{i=1}^{M} t_{p,i} \), the arrival rate \( \lambda \) is held constant at 15 UEs/sec, ensuring comparable results. Table~\ref{tab:proc_times} summarizes the impact of varying processing times on system performance. At lower processing times (10 ms and 30 ms), both monolithic and Open RAN remain stable, with monolithic RAN demonstrating lower utilization, queue length, and delay due to its higher service rate. As processing time increases to 50 ms, both systems experience higher delays, with Open RAN affected more significantly due to additional transmission delays. At high processing times (100 ms), both architectures become unstable as utilization exceeds 1, leading to unbounded queue growth.

\begin{table}[!t]
\centering
\caption{Performance Metrics for Various Processing Times (\(\lambda = 15\) UEs/sec).}
\label{tab:proc_times}
\begin{tabular}{|c|c|c|c|}
\hline
\textbf{Proc. Time} & \textbf{\(\mu\) (UEs/sec)} & \textbf{Queue (\(L_s\))} & \textbf{Delay (\(W\)) (ms)} \\ 
\textbf{(ms)} & M / O & M / O & M / O \\ \hline\hline

10  & 93.02 / 65.57  & 0.19 / 0.30   & 12.81 / 19.77 \\ \hline
30  & 32.52 / 28.37  & 0.85 / 1.13   & 57.07 / 74.79 \\ \hline
50  & 19.70 / 18.09  & 3.17 / 4.88   & 212.76 / 323.62 \\ \hline
100 & 9.92 / 9.50    & - / -         & - / - \\ \hline
\end{tabular}
\end{table}

\begin{figure}[!t]
\centering
\includegraphics[width=\columnwidth]{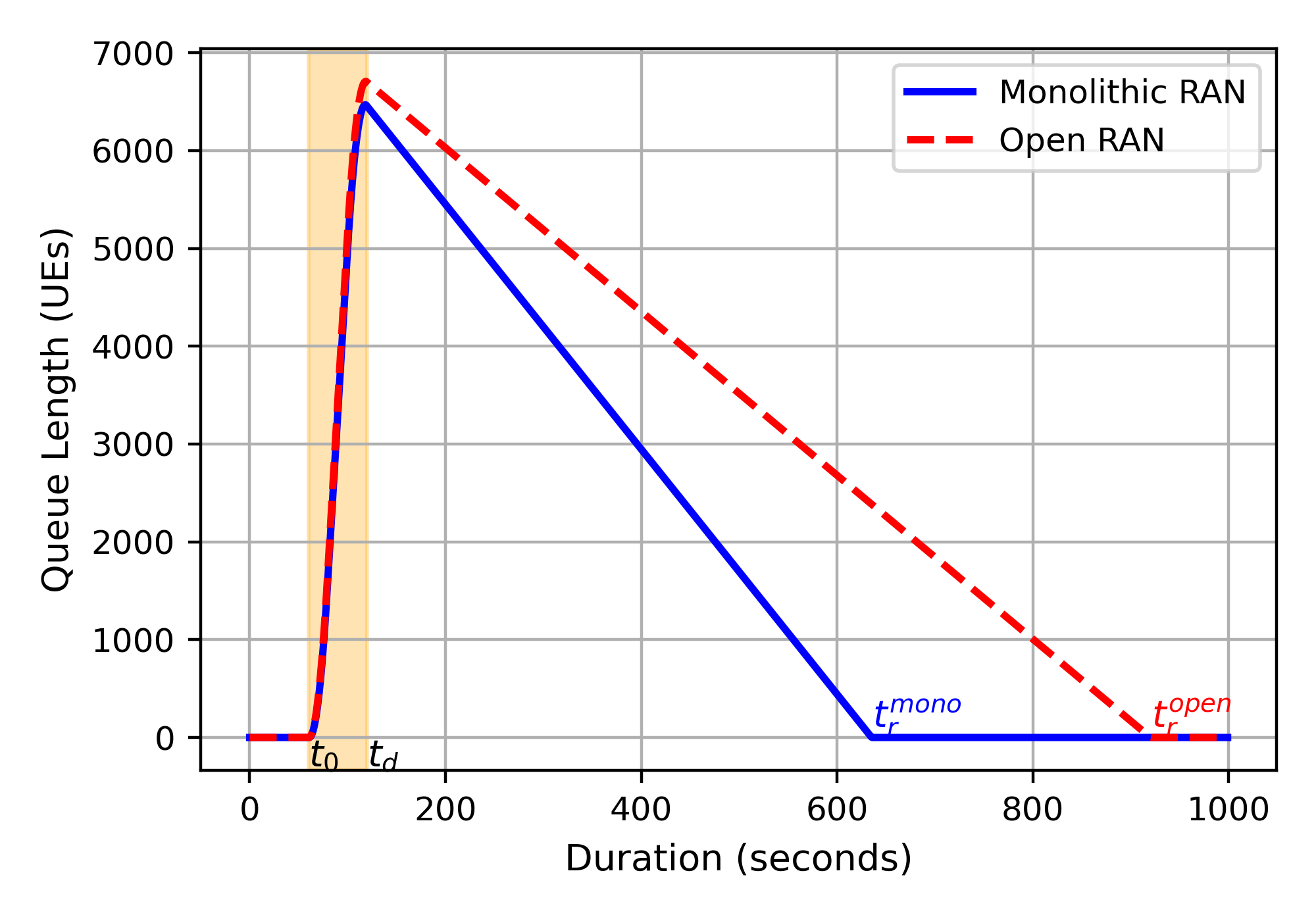}
\caption{Queue Length during Signaling Storm.}
\label{fig:ss_ql}
\end{figure}

\begin{figure}[!t]
\centering
\includegraphics[width=\columnwidth]{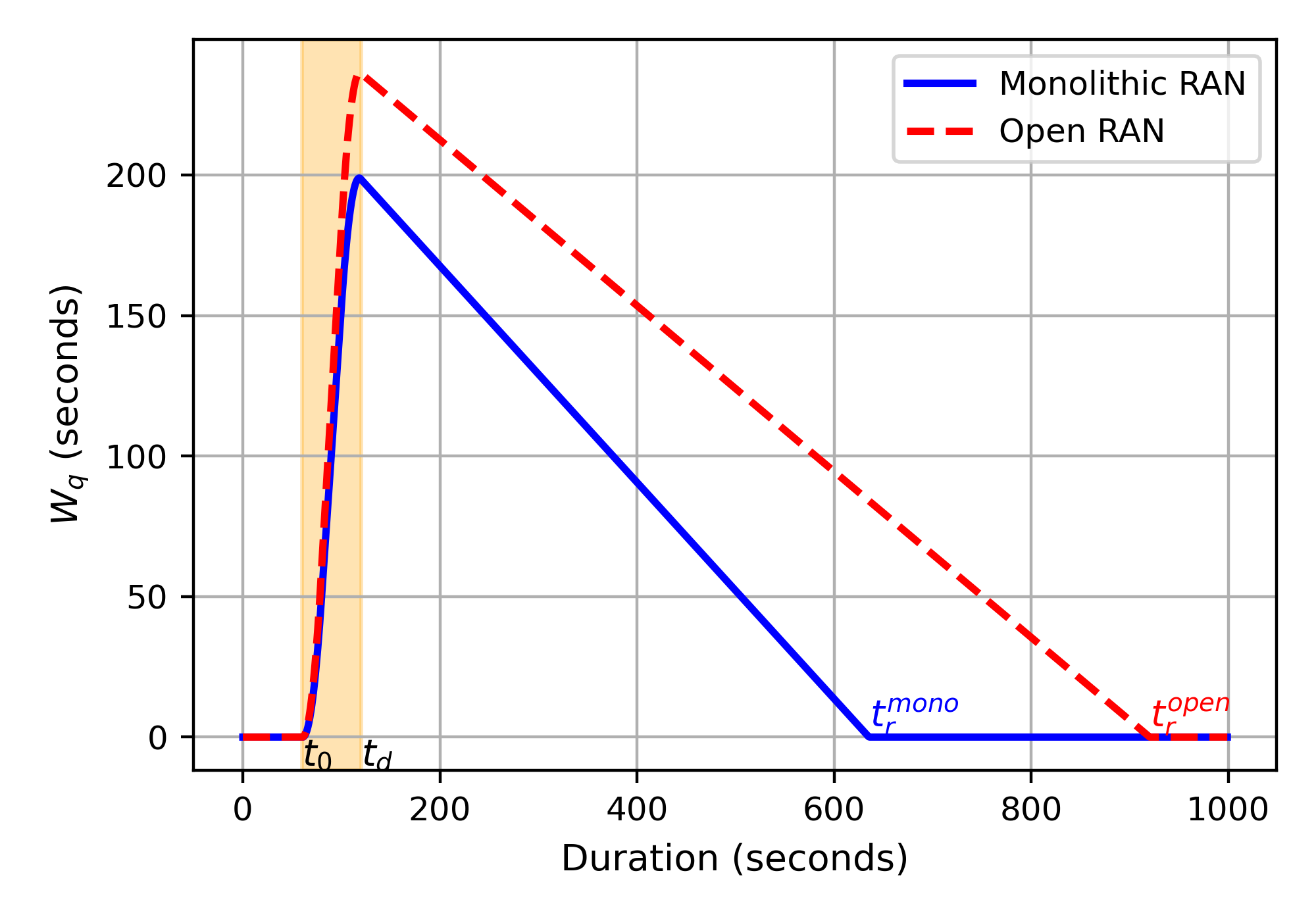}
\caption{Waiting Time of last UE during Signaling Storm.}
\label{fig:ss_wt}
\end{figure}

\begin{figure}[!t]
\centering
\includegraphics[width=\columnwidth]{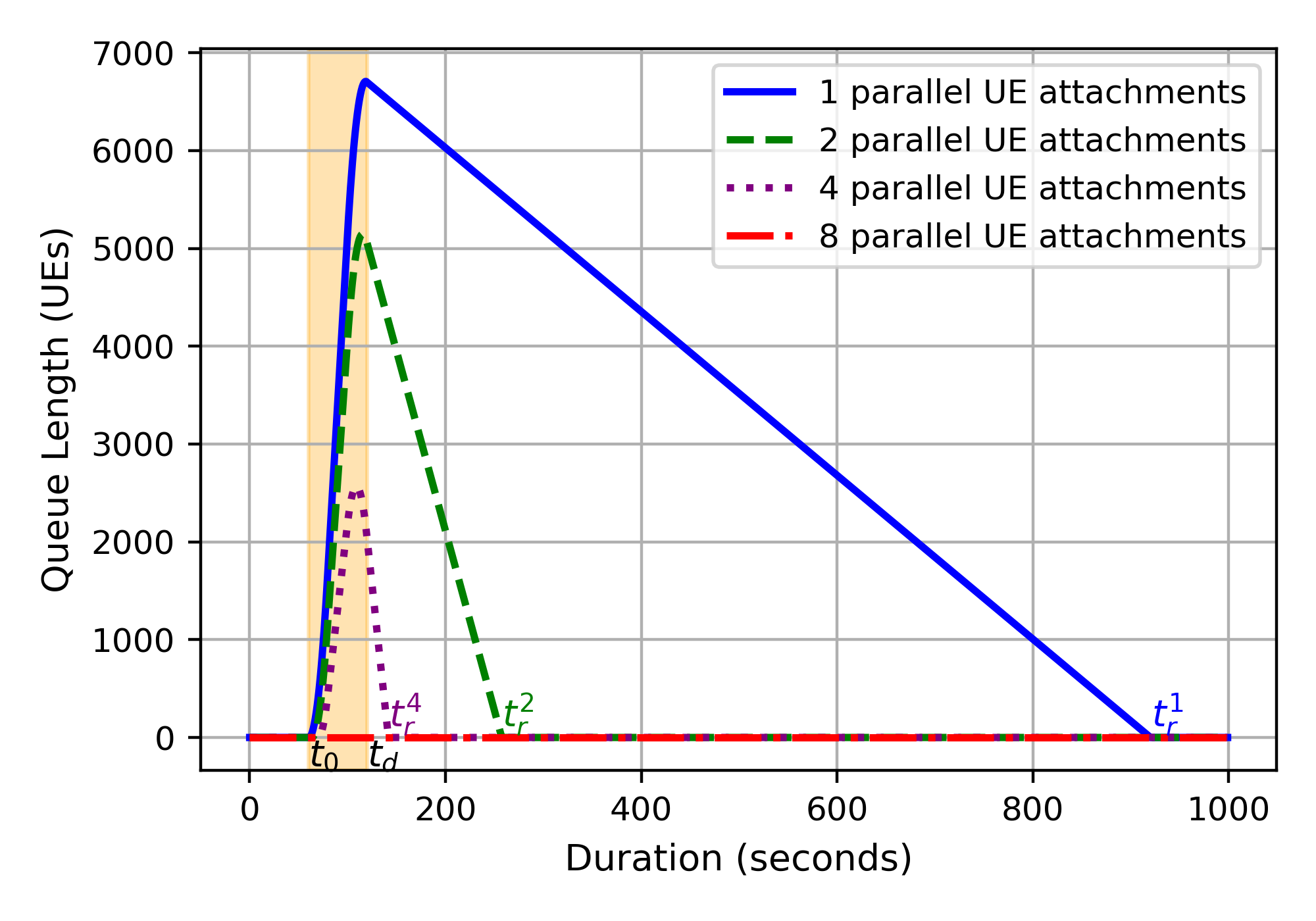}
\caption{Queue Length during Signaling Storm with Multiple Servers.}
\label{fig:ss_qlc}
\end{figure}

\begin{figure}[!t]
\centering
\includegraphics[width=\columnwidth]{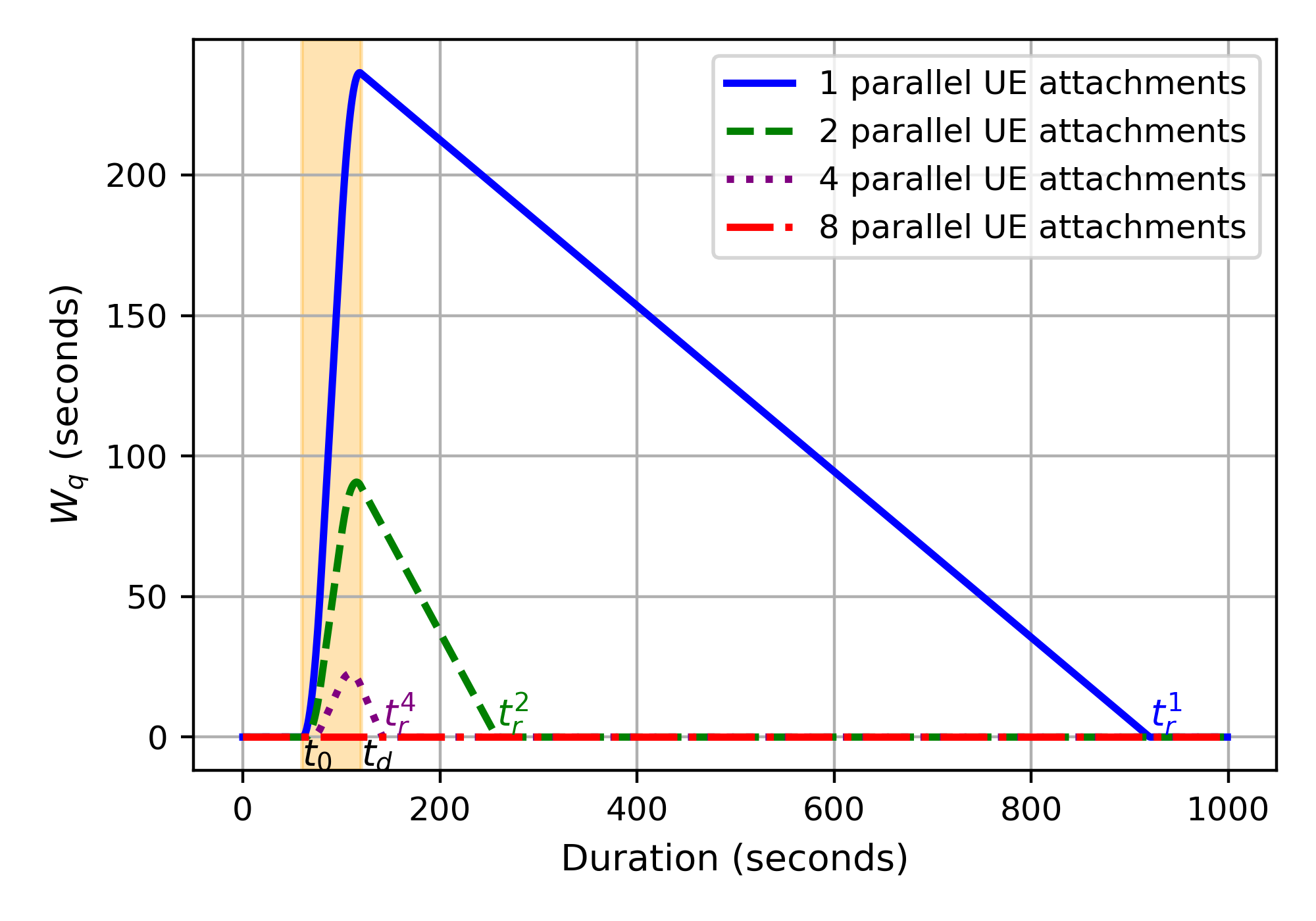}
\caption{Waiting Time of last UE during Signaling Storm with Multiple Servers.}
\label{fig:ss_wtc}
\end{figure}

To further understand system behavior under extreme conditions, we examine the impact of a signaling storm on queue length and waiting time, as shown in Figs.~\ref{fig:ss_ql} and~\ref{fig:ss_wt}. Open RAN experiences a higher and more prolonged queue buildup, indicating greater susceptibility to congestion under extreme loads. One approach to mitigating this congestion is the introduction of multiple servers, which helps reduce queue buildup and accelerate queue depletion, as shown in Figs.~\ref{fig:ss_qlc} and~\ref{fig:ss_wtc}. Increasing the number of servers lowers waiting times during peak congestion, thereby improving network resilience. The number of servers required to handle a storm with arrival rate \( \lambda_{\text{storm}} \) must satisfy \( c \geq \left\lceil \frac{\lambda_{\text{storm}}}{\mu} \right\rceil \).

\begin{figure}[!t]
    \centering
    \includegraphics[width=\columnwidth]{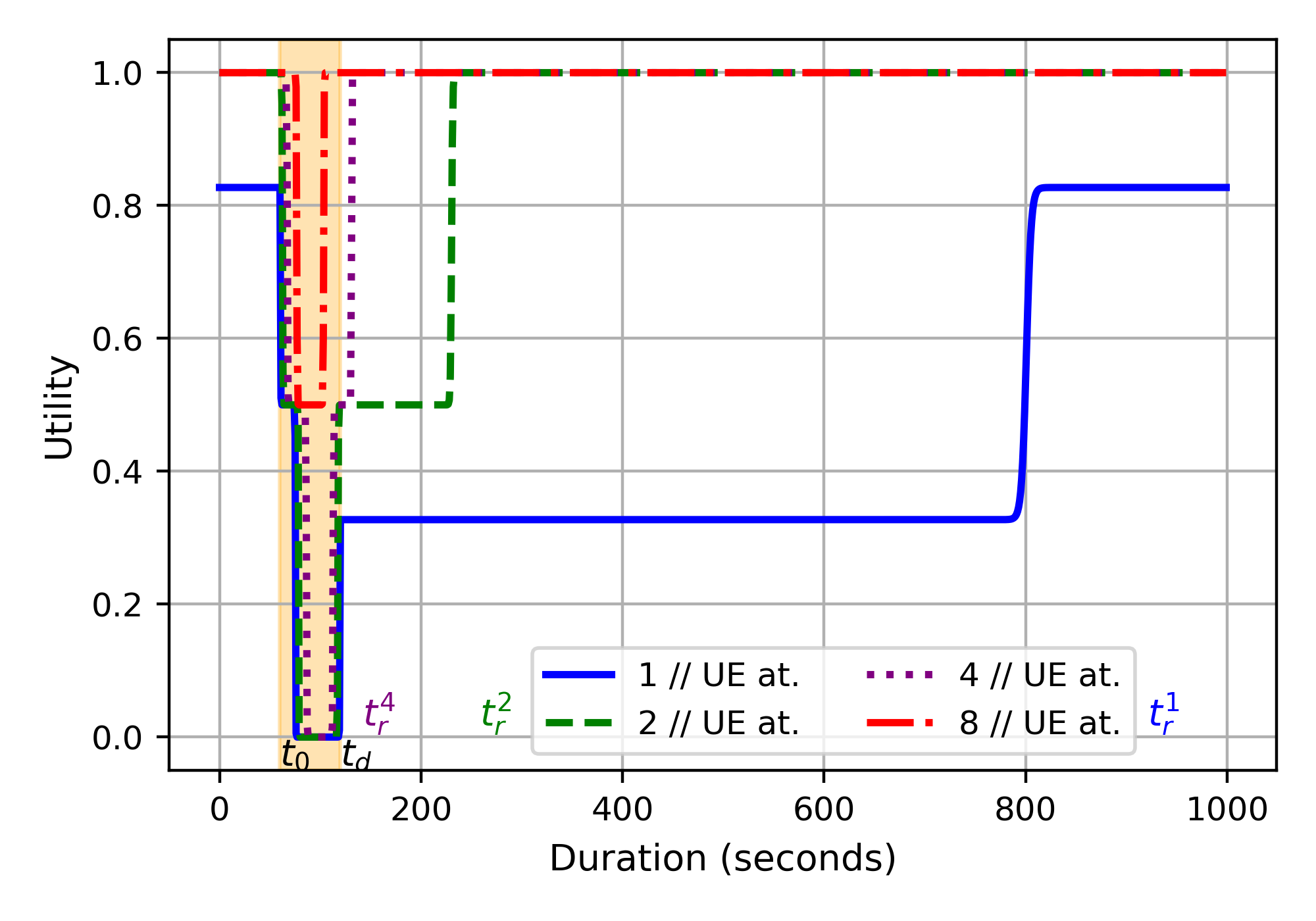}
    \caption{Utility vs. Duration for Different UEs.}
    \label{fig:utility_multiple_servers}
\end{figure}
Lastly, it is also important to assess the overall system stability during and after a signaling storm. Fig.~\ref{fig:utility_multiple_servers} illustrates the utility function \( u(t) \), showing two major drops: one due to sudden network load changes and another due to congestion. Once the storm subsides, the utility function gradually returns to its initial state, reflecting system recovery. To further evaluate the resilience of our system, we conducted a series of experiments comparing fixed server allocations with our Lyapunov-based adaptive mechanism. Our simulations modeled a signaling storm scenario with a stepwise increase and decrease in arrival rates, testing the system's ability to recover efficiently. We first analyzed the resilience of fixed server allocations using different values of $c$ (1, 2, 4, and 6) to determine how increasing the number of servers affects system performance. Then, we compared these results against the Lyapunov-based adaptation approach under different tuning parameters $V$ and $W$. Table~\ref{tab:resilience_comparison} presents a comparative analysis of resilience scores for fixed server allocations and adaptive mechanisms. The results indicate that lower fixed values of $c$ struggle to maintain resilience, with $c=1$ achieving a resilience score of only 0.197, significantly lower than the reference adaptive mechanism (0.762, a 286\% improvement). As $c$ increases, resilience improves, reaching 0.495 at $c=4$ and 1.000 at $c=6$. However, the Lyapunov-based adaptive mechanism consistently outperforms most fixed configurations, particularly when optimized with appropriate $V$ and $W$ values. For instance, the adaptive mechanism with $(V=1000, W=1)$ achieves a resilience score of 0.956, a 25.5\% increase over the reference adaptive case, demonstrating the benefits of dynamic resource allocation. These findings highlight the advantage of adaptive mechanisms in enhancing system recovery and stability compared to static configurations.

\begin{table}[!t]
\centering
\caption{Comparison of Fixed Number of Servers ($c$) and Adaptive Mechanism Resilience Scores.}
\label{tab:resilience_comparison}
\begin{tabular}{|c|c|c|c|c|}
\hline
$c$ & P & $V$ & $W$ \\
\hline \hline
1 & 0.197375 & -- & -- \\
2 & 0.340468  & -- & -- \\
4 & 0.495878 & -- & -- \\
6 & 1.000000  & -- & -- \\
Ref. Adapt. & 0.761993  & 1 & 1 \\
Adapt. & 0.758011  & 1 & 1000 \\
Adapt. & 0.956405  & 1000 & 1 \\
\hline
\end{tabular}
\end{table}

Building on these findings, enhancing resilience in Open RAN requires a combination of proactive and reactive mechanisms to mitigate the effects of signaling storms and improve overall network stability. The following strategies contribute to achieving resilience in such dynamic network environments:
\begin{itemize}
    \item \textbf{Traffic Monitoring and Anomaly Detection:} Proactively anticipating potential disruptions through real-time traffic monitoring and AI/ML-driven anomaly detection ensures early identification of abnormal patterns. This approach not only reduces the risk of signaling storms but also enhances network adaptability by dynamically retraining models to respond to evolving traffic anomalies. Improved network availability and security can be achieved by isolating potentially malicious UEs.

    \item \textbf{Dynamic Resource Scaling:} Absorbing the impact of high signaling loads is facilitated by dynamically allocating computing power, memory, and network resources based on real-time demand or predictive forecasts. This mechanism ensures consistent performance and minimizes service degradation during peak traffic or unexpected surges.

    \item \textbf{Fault Tolerance:} Adaptation mechanisms, such as redundancy and self-healing, are critical to maintaining functionality during failures. Virtualized network components equipped with backup instances ensure seamless service continuity, while automated fault detection and isolation minimize the scope and duration of disruptions.

    \item \textbf{Fast Recovery:} Recovery mechanisms leverage adaptive techniques such as reconfiguring network slices, reallocating resources, and prioritizing critical traffic flows to restore service levels quickly after disruptions. This ensures minimal downtime and continuous operation of mission-critical applications.
\end{itemize}

By adopting these proactive and reactive mechanisms and leveraging a well-defined utility and resilience metric, Open RAN networks can transition from robustness-focused designs to resilience-by-design frameworks. This approach ensures that networks can withstand disruptions, adapt dynamically, and recover efficiently, addressing the challenges posed by signaling storms and other adverse conditions.

\section{Conclusions}\label{sec:concl}
This study provides an in-depth analysis of Open RAN's disaggregated architecture, focusing on the additional overheads introduced during the UE initial attachment process.
By comparing Open RAN with traditional monolithic RAN, we examine the impact of disaggregation on UE attachment, showing that while both architectures experience increased delays and heightened risk of signaling storms as attachment rates rise, Open RAN's added overheads make it slightly more prone to congestion under high-load conditions.
Nevertheless, Open RAN’s architectural flexibility, enabled by virtualization, provides significant advantages in adapting to dynamic network conditions. This flexibility supports efficient resource allocation and the rapid deployment of congestion control mechanisms to mitigate signaling storms. To quantify and compare resilience in Open RAN deployments, we introduced a novel utility function that incorporates key factors such as UE arrival rates and queue length. Using this function, our results show that the proposed adaptive mechanism improves resilience by up to 286\% compared to fixed configurations, achieving resilience scores as high as 0.96 under optimal conditions. These findings highlight Open RAN’s potential to mitigate the effects of disaggregation and enhance network robustness under high-load conditions.

Future work will extend this analysis by leveraging an Open RAN emulator to further investigate the effects of disaggregation on signaling storms and validate the simulation findings in a more realistic setting. Additionally, we aim to explore a reinforcement learning-based approach to dynamically adjust the number of servers in response to signaling storm attacks. By leveraging adaptive decision-making, this method could enhance network resilience by optimizing resource allocation in real-time, ensuring efficient mitigation of extreme congestion events.

\section*{Acknowledgment}
This work has been supported by CHEDDAR: Communications Hub for Empowering Distributed Cloud Computing Applications and Research funded by the UK EPSRC under grant numbers EP/Y037421/1 and EP/X040518/1.

\bibliographystyle{IEEEtran}
\bibliography{references.bib}
\begin{IEEEbiography}[{\includegraphics[width=1in,height=1.25in,clip,keepaspectratio]{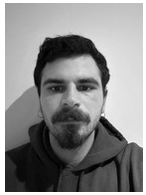}}]{Sotiris Chatzimiltis }
received a B.Sc. degree in computer science from the University of Cyprus, Cyprus, in 2021, and an M.Sc. degree in computer vision, machine learning and robotics from the University of Surrey, U.K., in 2022. He is a PhD student at the Institute for Communication Systems at the University of Surrey. His current research interests include machine learning, intrusion detection systems, distributed AI and Open RAN security.
\end{IEEEbiography}
\begin{IEEEbiography}
[{\includegraphics[width=1in,height=1in,clip,keepaspectratio]{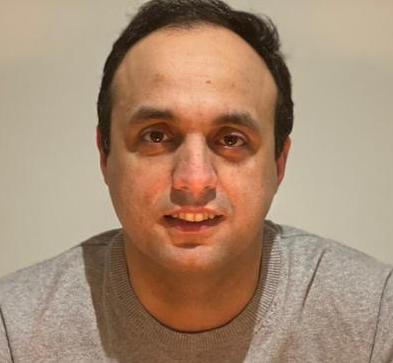}}]{Mohammad Shojafar} \textbf{(M'17-SM'19)} is an Associate Professor in network security and an Intel Innovator, professional ACM member and ACM distinguished speaker, a fellow of the Higher Education Academy, and a Marie Curie alumnus, working in the 5G \& 6G Innovation Centre (5G/6GIC), Institute for Communication Systems (ICS), at the University of Surrey, UK. Before joining 5G/6GIC, he was a senior researcher and a Marie Curie fellow in the SPRITZ Security and Privacy Research group at the University of Padua, Italy. Dr Mohammad secured around $\pounds$1.9M as PI in various EU/UK projects, including ORAN-TWIN (funded by EPSRC/DSIT CHEDDAR Hub UK;2024), D-XPERT (funded by I-UK/UK;2024), 5G MoDE (funded by DSIT/UK;2023), 5G ONE4HDD (funded by DSIT/UK;2023),  TRACE-V2X (funded by EU/MSCA-SE;2023), AUTOTRUST (funded by ESA/EU;2021), PRISENODE (funded by EU/MSCA-IF:2019), and SDN-Sec (funded by Italian Government:2018). He was also COI of various UK/EU projects like HiPER-RAN (funded by DSIT/UK;2023), APTd5G project (funded by EPSRC/UKI-FNI:2022), ESKMARALD (funded by UK/NCSC;2022), GAUChO, S2C and SAMMClouds (funded by Italian Government;2016-2018). He received his PhD in ICT from Sapienza University of Rome, Rome, Italy, in 2016 with an ``Excellent'' degree. He is an associate editor in \textit{IEEE Transactions on Network and Service Management}, \textit{IEEE Transactions on Intelligent Transportation Systems}, \textit{IEEE Transactions on Green Communications and Networking}, \textit{IEEE Transactions on Consumer Electronics}, and Computer Networks. For additional information:  
\url{https://www.surrey.ac.uk/people/mohammad-shojafar}.
\end{IEEEbiography}
\begin{IEEEbiography}[{\includegraphics[width=1in,height=1.25in,clip,keepaspectratio]{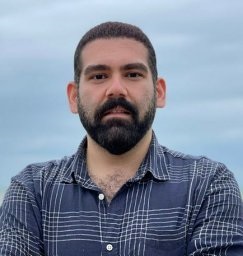}}]{Mahdi Boloursaz Mashhadi}\textbf{(S’14-M’18, SM’23)} is a Lecturer at the 5G/6G Innovation Centre (5G/6GIC) at the Institute for Communication Systems (ICS), University of Surrey (UoS), and a Surrey AI fellow. His research is focused at the intersection of AI/ML with wireless communication, learning and communication co-design, generative AI for telecommunications, and collaborative machine learning. He received B.S., M.S., and Ph.D. degrees in mobile telecommunications from the Sharif University of Technology (SUT), Tehran, Iran. He has more than 40 peer reviewed publications and patents in the areas of wireless communications, machine learning, and signal processing. He is a PI/Co-PI for various government and industry funded projects including the UKTIN/DSIT 12M£ national project TUDOR. He received the Best Paper Award from the IEEE EWDTS conference, and the Exemplary Reviewer Award from the IEEE ComSoc
in 2021 and 2022. He served as a panel judge for the International Telecommunication Union (ITU) on the “AI/ML in 5G” challenge 2021-2022. He is an associate editor for the Springer Nature Wireless Personal Communications Journal.
\end{IEEEbiography}
\begin{IEEEbiography}[{\includegraphics[width=1in,height=1.25in,clip,keepaspectratio]{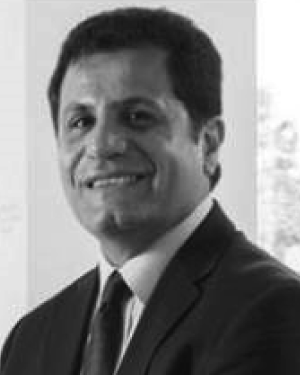}}]{Rahim Tafazolli }\textbf{(Fellow, IEEE)} is currently a Professor in mobile and personal communications and the Director of the Institute of Communication Systems (ICS) and the 5G and 6G Innovation Centre, University of Surrey. He has been active in research for over 30 years and published more than 1200 research papers. He has been a technical advisor to many mobile companies, and has lectured, chaired, and been invited as keynote speaker to a number of IEE and IEEE workshops and conferences. He served as the Chairperson for the EU Expert Group on Mobile Platform (e-mobility SRA), the Chairperson for the Post-IP working Group in e-mobility, and the past Chairperson of WG3 of WWRF. He is nationally and internationally known in the field of mobile communications. In May 2018, he was appointed as a Regius Professor in electronic engineering for recognition of his exceptional contributions to digital communications technologies over the past 30 years. He was elected as a fellow of the U.K. Royal Academy of Engineering, in 2020. He is a fellow of IET and the Wireless World Research Forum (WWRF).
\end{IEEEbiography}
\end{document}